\documentclass[sigplan,screen]{acmart}
\usepackage[labelfont=bf]{caption}
\usepackage{subcaption}
\usepackage[font={small,it}]{caption}
\usepackage{color}
\usepackage{tikz}
\usepackage{amsmath}
\usepackage{subcaption}
\usepackage{graphicx}
\usepackage{tabularx}
\usepackage{array,multirow,rotating}
\usepackage{algorithm}
\usepackage{algpseudocode}
\usepackage{amsmath}
\usepackage{amssymb}
\usepackage{tikz}
\usepackage{hyperref}
\usepackage{pifont}
\newcommand{\cmark}{\ding{51}}%
\newcommand{\xmark}{\ding{55}}%

\usepackage[compact]{titlesec}
\titlespacing\section{0pt}{4pt plus 0pt minus 1pt}{2pt plus 0pt minus 1pt}
\titlespacing\subsection{0pt}{4pt plus 0pt minus 1pt}{1pt plus 0pt minus 1pt}

\usepackage[inline]{enumitem}
\newcommand*\circled[1]{\tikz[baseline=(char.base)]{
          \node[shape=circle,draw,fill=black,text=white,font=\bf,inner sep=0.5pt] (char)
            {\scriptsize#1};}}
            
\newcommand*\tcircled[1]{\tikz[baseline=(char.base)]{
            \node[shape=circle,draw,thick,font=\bf,inner sep=1pt] (char) 
            {\footnotesize#1};}}
            
\usepackage{wrapfig}
\usepackage{soul}

\AtBeginDocument{%
  \providecommand\BibTeX{{%
    \normalfont B\kern-0.5em{\scshape i\kern-0.25em b}\kern-0.8em\TeX}}}

\begin{abstract}
Datacenters are witnessing a rapid surge in the adoption of serverless functions for microservices-based applications. A vast majority of these microservices typically span less than a second, have strict SLO requirements, and are chained together as per the requirements of an application. The aforementioned characteristics introduce a new set of challenges, especially in terms of container provisioning and management, as the state-of-the-art resource management frameworks, employed in serverless platforms, tend to look at microservice-based applications similar to conventional monolithic applications. Hence, these frameworks suffer from microservice-agnostic scheduling and colossal container over-provisioning, especially during workload fluctuations, thereby resulting in poor resource utilization.

In this work, we quantify the above shortcomings using a variety of workloads on a multi-node cluster managed by Kubernetes and Brigade serverless framework. To address them, we propose \emph{Fifer} --- an adaptive resource management framework to efficiently manage function-chains on serverless platforms. The key idea is to make \emph{Fifer} (i) utilization conscious by efficiently bin packing jobs to fewer containers using function-aware container scaling and intelligent request batching, and  (ii) at the same time, SLO-compliant by proactively spawning containers to avoid cold-starts, thus minimizing the overall response latency. Combining these benefits, \emph{Fifer} improves container utilization and cluster-wide energy consumption by 4$\times$ and 31\%, respectively, without compromising on SLO's,  when compared to the state-of-the-art schedulers employed by serverless platforms. 

\end{abstract}

\renewcommand\footnotetextcopyrightpermission[1]{}
 \sloppypar

\def\footnoterule{\kern-3pt
  \hrule \kern 2.6pt} 
\begin{document}

\title{Fifer: Tackling Underutilization in the Serverless Era}
\author{Jashwant Raj Gunasekaran}
\affiliation{%
 \institution{Penn State}}
 \author{Prashanth Thinakaran}
\affiliation{%
 \institution{Penn State}}
 \author{Nachiappan Chidambaram}
\affiliation{%
 \institution{Penn State}}
 \author{Mahmut T. Kandemir}
\affiliation{%
 \institution{Penn State}}
 \author{Chita R. Das}
\affiliation{%
 \institution{Penn State}}

\renewcommand{\shortauthors}{Pre-Print Version}



\pagestyle{fancy}
\maketitle

\section{Introduction} 
\label{sec:intro}

The advent of public clouds in the last decade has led to the explosion in the use of microservice-based applications~\cite{gan2019open}. 
%
Large cloud-based companies like Amazon~\cite{aws}, Facebook~\cite{facebook}, Twitter~\cite{Twitter}, and Netflix~\cite{netflix} have capitalized on the ease of scalability and development offered by microservices, embracing it as a first-class application model~\cite{7515686}. For instance, a wide range of Machine Learning (ML) applications such as facial  recognition~\cite{bartlett2005recognizing}, virtual systems~\cite{sirius}, content recommendation~\cite{hazelwood2018applied}, etc., are realized as a series of inter-linked microservices\footnote{A microservice is the smallest granularity of an application performing an independent function, a.k.a functions in serverless domain.}, also known as \textit{microservice-chains}~\cite{wechat,8486300}.
These applications are user-facing~\cite{8675201} and hence, demand a strict service-level objective (SLO), which is usually under \texttt{1000 ms}~\cite{swayam,p1SLO,p3SLO}. It is, therefore, imperative to mitigate the end-to-end latency of a microservice-chain to provide a satisfactory user experience. The SLOs for such microservices are bounded by two factors -- (i) resource provisioning latency, and (ii) application execution time. As a majority of these microservices usually execute within a few milliseconds~\cite{djinn,sirius}, \emph{serverless functions} ~\cite{lambda,azure,ibm} have proven to be an ideal alternative over virtual machines (VM), as they not only have very short resource-provisioning latencies, but also abstract away the need for the users to explicitly manage the resources.

However, adopting serverless functions introduce a new set of challenges in terms of scheduling and resource management (RM) for the cloud providers~\cite{spec,DBLP:conf/socc/KaffesYK19}, especially when deploying large number of millisecond-scale function chains\footnote{We refer to microservice-chains and function chains interchangeably throughout the paper. Also, we refer to each function within a function-chain as a stage.}. There has been considerable prior work~\cite{cloudburst,openwhisk,akkus2018sand,suresh2019fnsched,234839,gg} in RM frameworks to leverage the asynchronicity, event-drivenness and scalability of serverless applications. 
Despite having these sophisticated frameworks, the resource management for thousands of short-lived function-chains still have significant inefficiencies in terms of resource utilization and SLO-compliance. Viewing functions in a function-chain as a truly independent entity, in fact, accentuates these inefficiencies.
Studying state-of-the-art RM frameworks, we identify three critical reasons for these inefficiencies.

$\bullet$ Most frameworks are built just to meet each individual function's SLOs. {\emph{Being imperceptive to the total end-to-end SLO of the function-chain}} leads to sub-optimal uniform scaling of containers at every function stage. This inherently leads to over-provisioning containers, which in turn results in large number of machines to host idle containers thereby increasing the provider's operating costs.

$\bullet$ Many frameworks {\emph {employ one-to-one mapping of requests to containers}}~\cite{peeking}. 
This inherently leads to excessive number of containers being provisioned when handling a sudden burst of requests than that are actually needed to meet the application-level SLOs. 

$\bullet$ Lastly, in the quest to reduce the number of provisioned containers, certain frameworks ~\cite{azure,google} make use of {\emph{naive batching of requests on to a static pool of containers}}. Fixing the number of containers in an application agnostic manner results in SLO violations, especially for functions with strict SLO requirements.     

These inefficiencies collectively open the door towards stage-aware resource management by exploiting the ``leftover slack'' in these function chains. Leveraging slack allows individual functions to be queued in batches at existing containers without violating the application-level SLOs. In this paper, we present, \textit{Fifer}, which to the best of our knowledge, is the first work that employs stage-aware container provisioning and management of function chains for serverless platforms. \emph{Fifer}~\footnote{A \emph{Fifer} plays a small flute to help soldiers in a brigade (or battalion) to keep their marching pace in coordination with the drummers. In spirit, our framework helps the Brigade system in Kubernetes to adapt to functions-chains by being proactive and stage-aware.} makes use of novel slack estimation techniques for individual functions and leverages it to significantly reduce the number of containers used, thereby leading to increased resource utilization and cluster energy efficiency. While slack-based request batching can significantly minimize the number of containers spawned, it still leads to SLO violations because of cold-starts, especially during dynamic load fluctuations. \emph{Fifer} makes use of proactive container provisioning using load prediction models to minimize the SLO violations incurred due to cold starts. To this end, the \textbf{key contributions} of the paper are the following:

$\bullet$ We {\em characterize the effect of cold-starts} for various ML inference applications on AWS serverless platforms and show that they have a large disparity in container provisioning times compared to application execution times. Further, we show that for an incoming series of requests, queuing them for batched execution at warm containers can greatly reduce the number of containers being spawned. 

$\bullet$ 
We introduce the notion of slack, which is defined as the difference between execution time and overall response latency. 
We propose \textit{Fifer}, which takes advantage of this slack towards calculating the batch-size to determine the optimal number of requests that could be grouped at every stage.
\textit{Fifer} is inherently \emph{stage aware}, such that it allocates slack to every function stage of an application proportionate to its execution time, and independently decides the scale-out threshold for every stage. 

$\bullet$ We quantitatively characterize the benefits of using different load prediction models (ML and non-ML) to enable proactive container provisioning. Based on our findings, we implement \emph{Fifer} with a novel \emph{LSTM-based~\cite{lstm} prediction model}, which provides fairly accurate request arrival estimations even when there are large dynamic variations in the arrival rate. 

$\bullet$ We implement \emph{Fifer} as a part of the Brigade serverless workflow framework~\cite{brigade} in a Kubernetes cluster and extensively evaluate it with different request arrival patterns using both synthetic traces and comprehensive real-world traces to show its advantage over other frameworks. Our results from the experimental analysis on an 80 core cluster and extensive large-scale simulations show that \emph{Fifer} spawns up to 80$\%$ fewer containers on an average, thereby improving container utilization and cluster-wide energy savings by up to 4$\times$\ and 31\%, respectively, when compared to state-of-the art non-batching reactive schedulers employed in serverless platforms. 

\section{Background and Motivation} 
\label{sec:motivation}
We start with providing a brief overview of serverless function-chains followed by a detailed analysis of their performance to motivate the
need for \emph{Fifer}.
\subsection{Serverless Functions Chains}
\label{whyserverless}
\begin{figure}
\centering
    \includegraphics[width=0.4\textwidth]{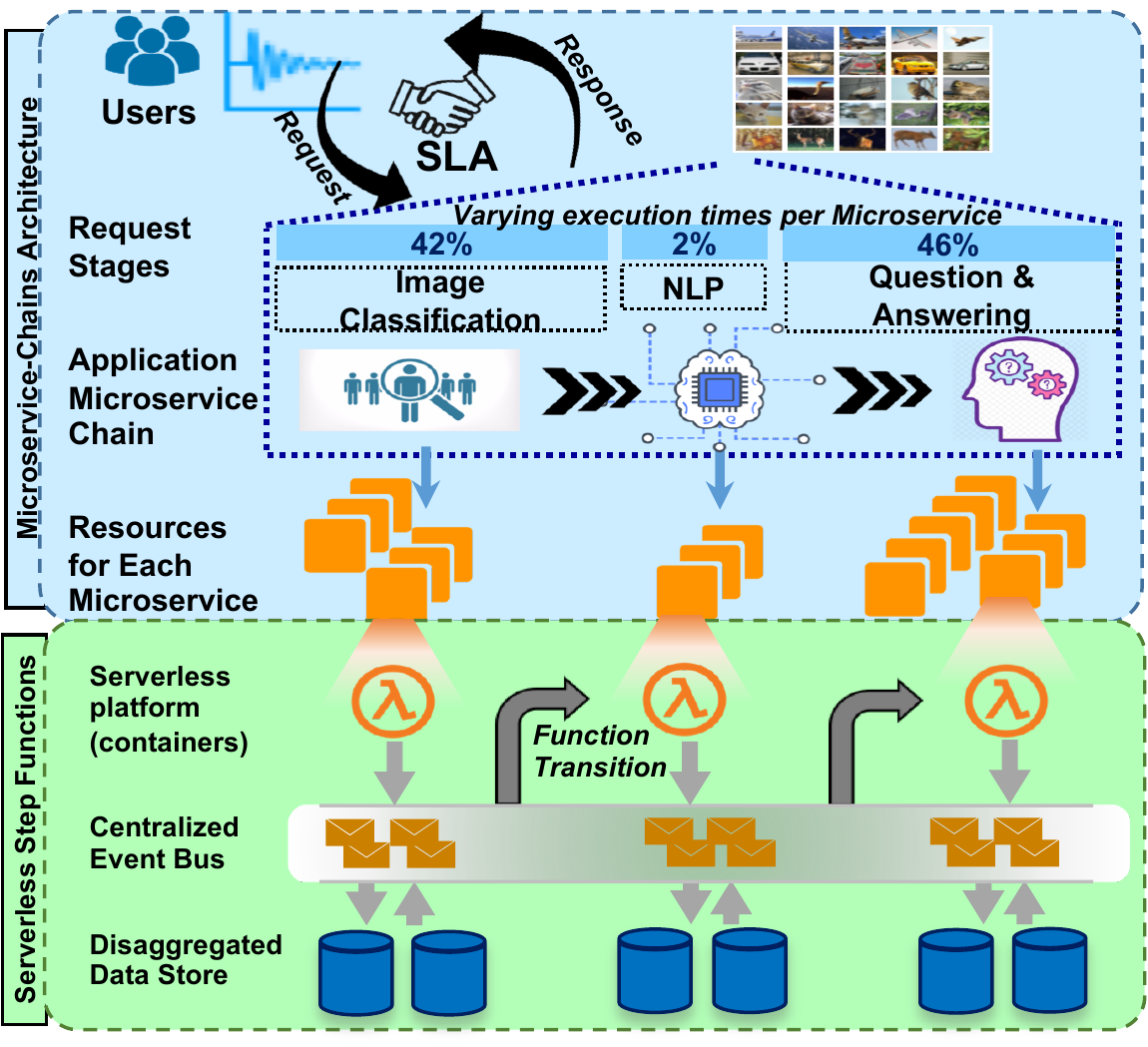}
    \caption{The blue and green box show the architectures of a typical microservice-chain and a serverless function respectively.}
    \label{step-functions}
\end{figure}

The overall framework for deploying microservice-chains in serverless platforms is shown in Figure~\ref{step-functions}. Multiple serverless functions (with one function per microservice) are stitched together using synchronization services such as AWS Step Functions~\cite{aws-step,durable,composer,gcp-dataflow} to form a ``function-chain''. Though the whole function-chain can also be deployed as one monolithic function, splitting them has several known advantages, in terms of ease of deployment and scalability per microservice. The actual transition between each function pair is in the form of communication events over a centralized event bus. Due to the stateless natures of serverless functions, input data such as pre-trained models, etc., need to be retrieved from ephemeral data storage like AWS S3~\cite{s3}. For further details on serverless functions, we refer the reader to prior works~\cite{peeking,spec,openlambda,Jonas:2017:OCD:3127479.3128601,hellerstein2018serverless,faas,faas1,234835,feng2018exploring,faas2017report}. 
In the context of this paper, we specifically focus on scenarios where tenants choose serverless platforms to host their applications. These applications will in turn be used by multiple end-users. In the case of multi-tenancy, our proposed ideas can be individually applied to each tenant. Also, we limit our scope to container provisioning and management, and we do not address function communication and storage related bottlenecks.
\subsection{Shortcomings of Current Serverless Platforms}
\label{sec:coldstart}                  
We start by describing the two major implications observed in current serverless platforms, with respect to hosting individual functions and function-chains.
\subsubsection{Cold-Start Latency for Single Functions}         
\quad In serverless computing, when a function is invoked as a part of deploying the tenant application, it is launched within a newly-created container, which incurs a start-up latency known as cold-start. Though modern virtualization technologies like microVMs~\cite{firecracker} reduce container start-up time, majority of cold-start time is dominated by application and runtime initialization. To avoid cold-starts, public cloud providers like Amazon try to launch every function in warm containers (existing containers)~\cite{peeking} whenever available. However, if all warm containers are occupied, a new container has to be spawned, which usually takes a few seconds. For applications which execute within a few milliseconds, it is clear that this penalty would be significantly higher, especially when the applications are user-facing, where it is crucial to ensure satisfactory response time. 


\begin{figure}
\begin{minipage}[t]{0.99\linewidth}
\centering
\begin{subfigure}[t]{.49\linewidth}
\includegraphics[width=0.98\textwidth]{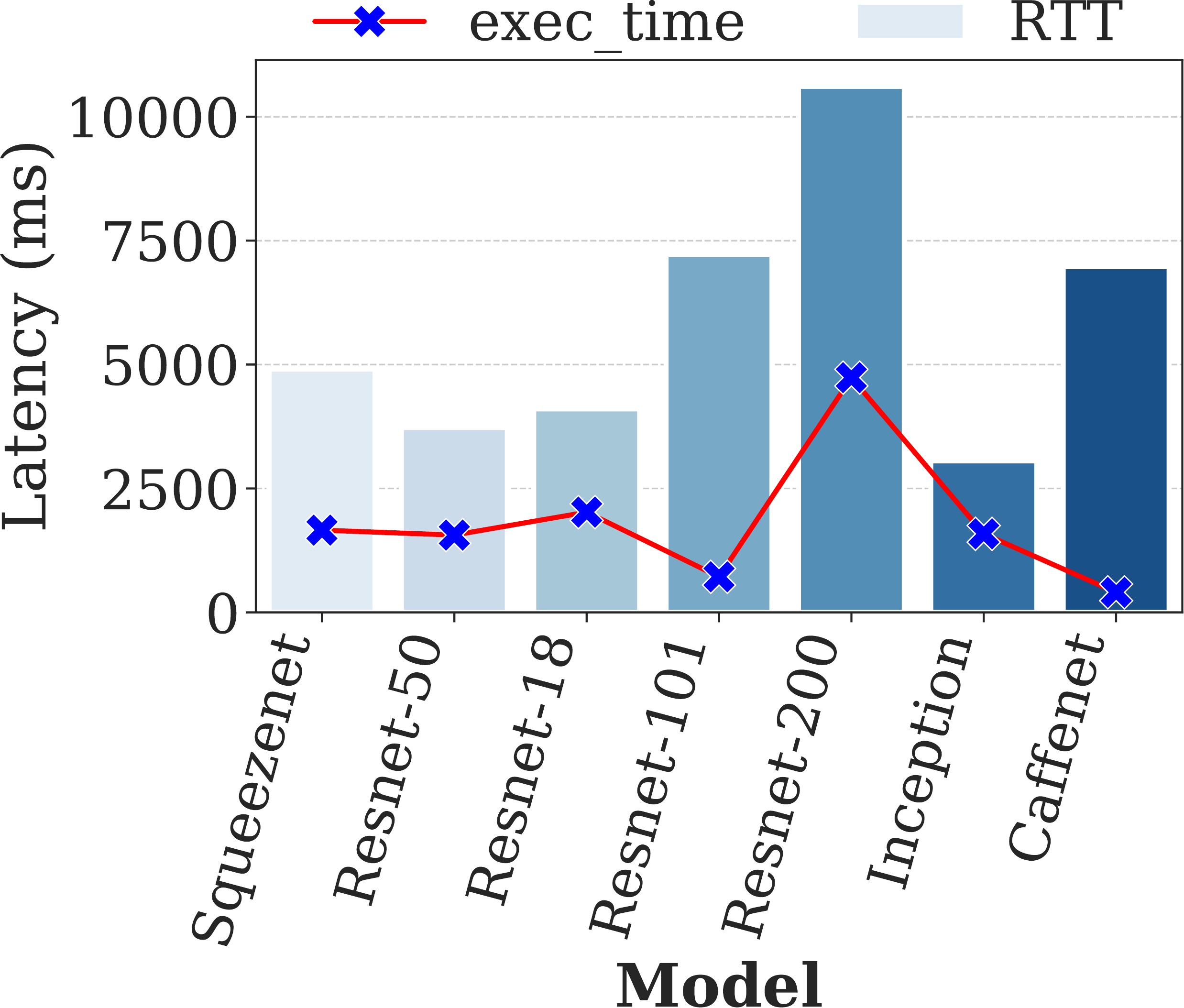}
\caption{Cold Start Latency.}
\label{model-startup}
\end{subfigure}
\begin{subfigure}[t]{.49\linewidth}
\includegraphics[width=0.98\textwidth]{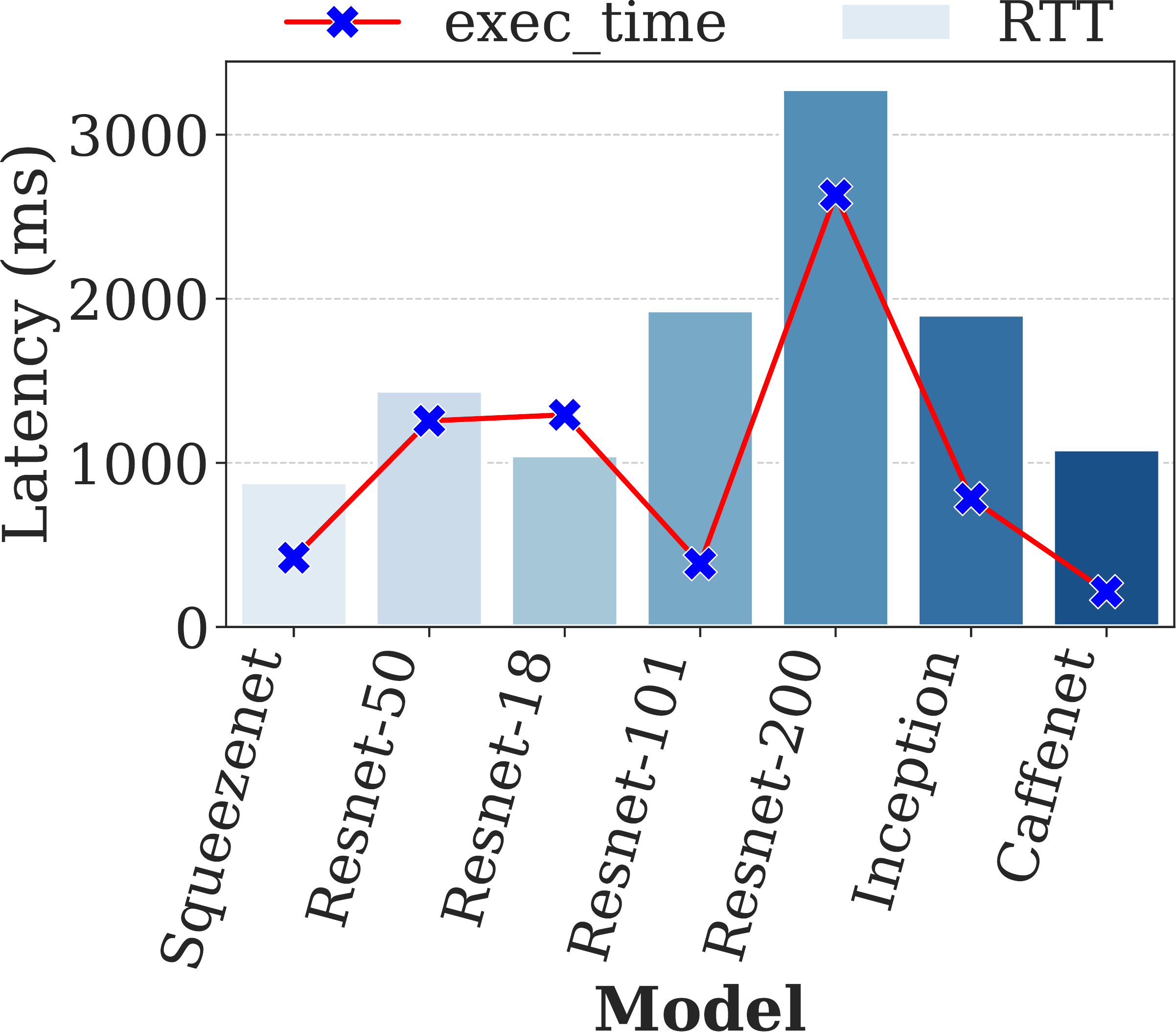}
\caption{Warm Start Latency.}
\label{rtt}
\end{subfigure}
\end{minipage} 
\caption{Implications of cold-starts for image inference on AWS serverless functions.}
\label{fig:startup}
\end{figure}   

To characterize the cold-start and warm-start latencies, we execute an ML image inference application using the \emph{Mxnet}~\cite{mxnet} framework on AWS lambda~\cite{lambda}. We use 7 different pre-trained ML models with varying execution times proportionate to the model sizes. Figure~\ref{fig:startup} plots the breakdown of total execution time for both cold and warm start as follows: (i) the time reported by AWS lambda for executing the inference (\texttt{exec\_time}), and (ii) the round-trip time (\texttt{RTT}) starting from the query submission at a client to receiving the response from AWS lambda. Cold start latency is measured for the first invocation of a request because there would be no existing containers to serve the request. Warm start is measured as an average latency of 100 subsequent requests, over a 5 minute interval. From Figure~\ref{model-startup}, it is evident that the cold start overheads on many occasions are higher than the actual query execution time, especially for larger models like \texttt{Resnet-200}. For warm starts, as shown in Figure~\ref{rtt}, the total time taken is within 1500 ms, except for larger models. From this, we can infer that the cold starts contribute {\raise.17ex\hbox{$\scriptstyle\sim$}}2000 to 7500 ms on top of execution time of the function itself. 

To avoid cold-starts, certain frameworks~\cite{fission, composer} employ a pre-warmed pool of idle containers which results in wasted memory consumption, in turn leading to energy inefficiency. This inefficiency can be potentially avoided for millisecond-scale applications (e.g., \texttt{Squeezenet}~\cite{squeeze} in Figure~\ref{fig:startup}) by allowing the requests to queue up on existing containers rather than launching them on separate ones. This can be done when the delay incurred from cold-starts is higher than the delay incurred from queuing the requests. Hence, the decision to queue versus starting up a new container depends on the SLO, execution times of the application and the cold-start latencies of the containers.
In contrast, RM frameworks used in Azure are known queue the incoming requests~\cite{peeking} on a fixed number of containers. Fixing the number of containers in an application agnostic manner will result in SLO violations, especially for functions with strict response latencies. 
\\ 
\textbf{Key takeaway:} \textit{Based on SLOs, cold-start latencies and execution times of applications, queuing functions can minimize the number of containers spawned without violating SLOs.}
\subsubsection{What is different with function-chains?}\label{sec:exec}
\quad In the case of function-chains consisting of a series of serverless functions (as described in Section \ref{whyserverless}), containers would be spawned individually for every stage. In existing serverless platforms, the RM framework would uniformly spawn containers at every stage depending on the request arrival load. However, the execution times of the functions at each stage are not uniform. 
\begin{figure}
\begin{minipage}[t]{0.99\linewidth}
\centering
\begin{subfigure}[t]{.48\linewidth}
    \includegraphics[width=0.95\textwidth]{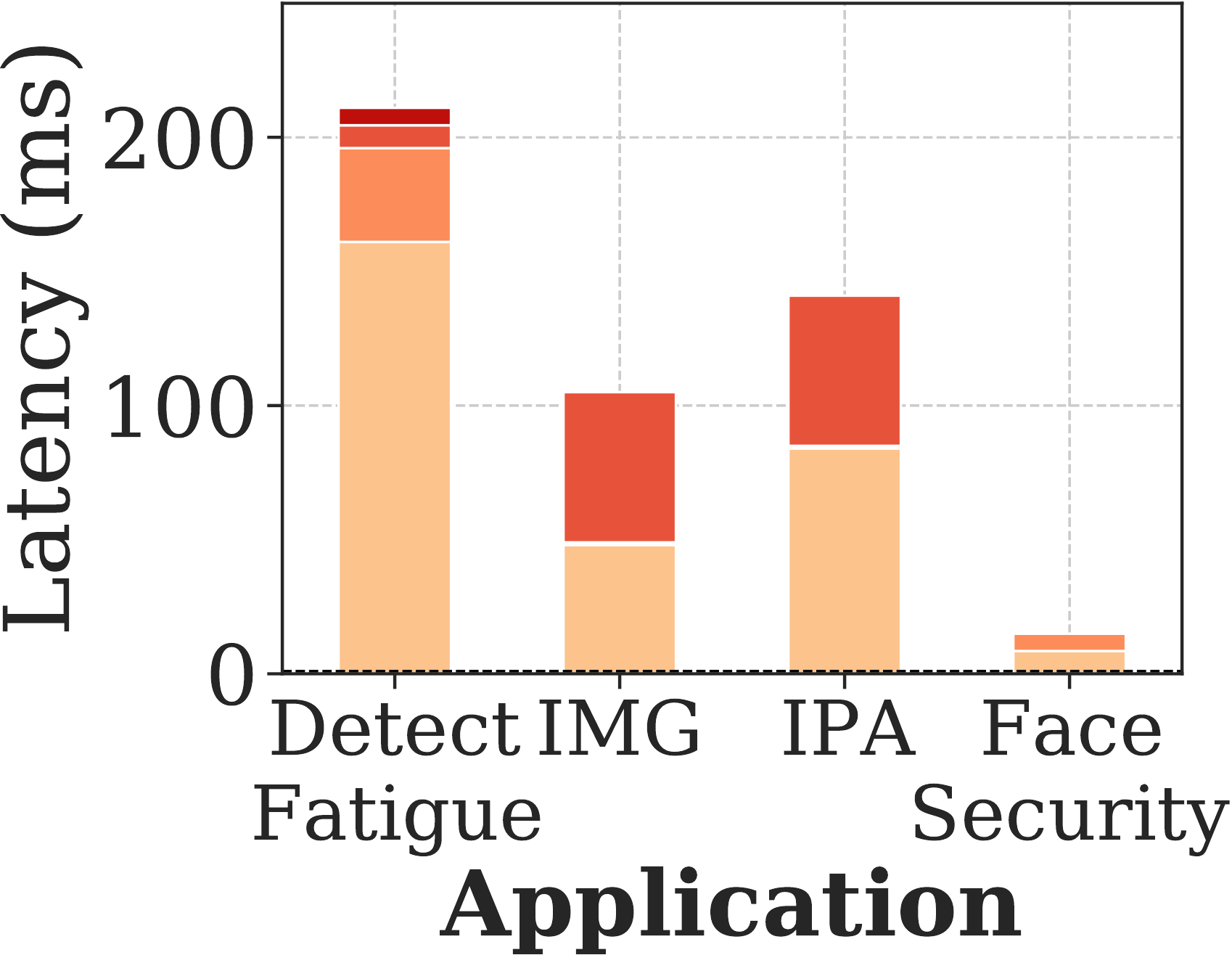}
\caption{Per-stage breakdown of overall application execution times.}
\label{fig:slack}
\end{subfigure}
\begin{subfigure}[t]{.48\linewidth}
\includegraphics[width=0.80\textwidth]{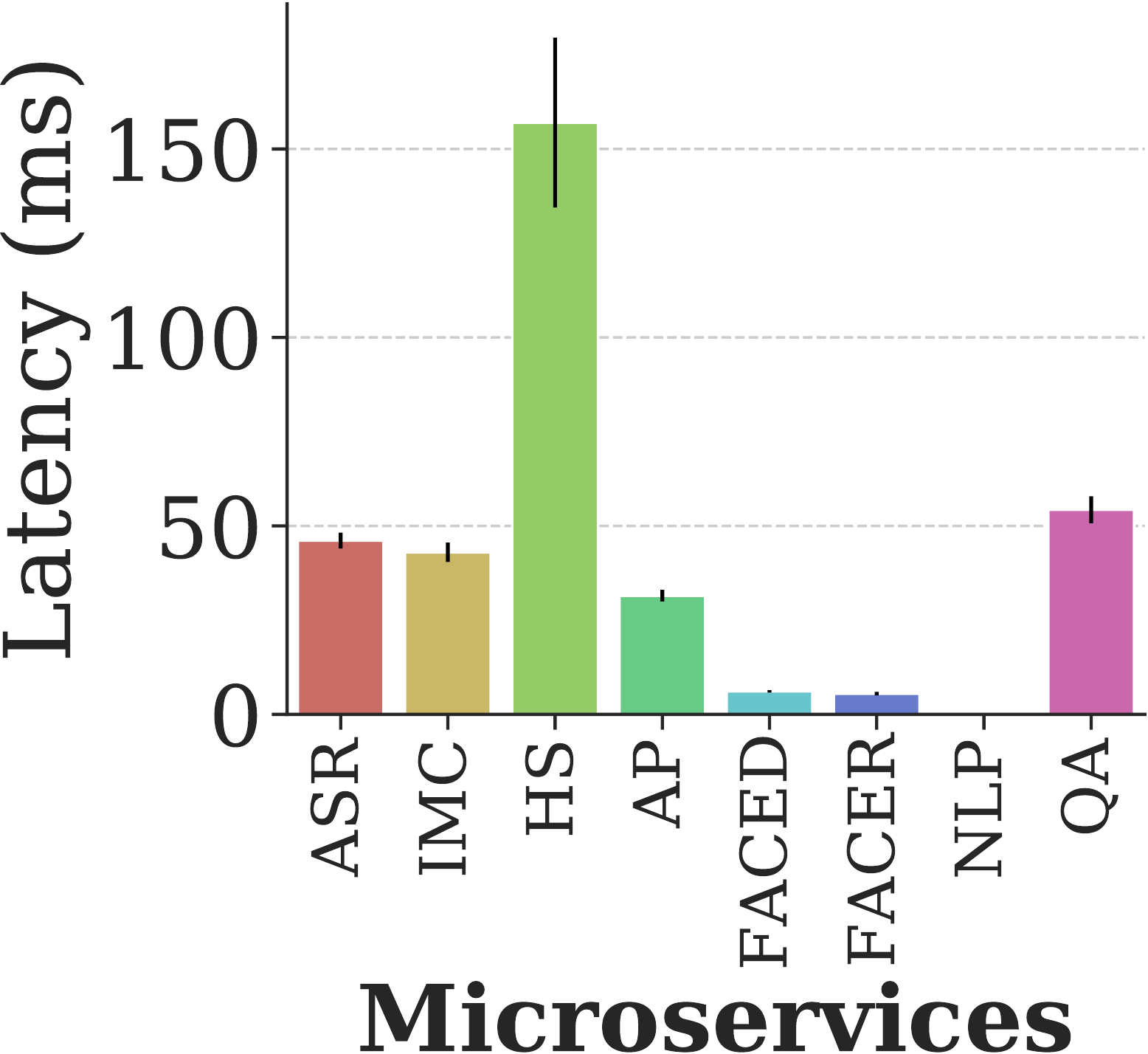}
\caption{Variation of execution time for each microservice.}
\label{fig:exec}
\end{subfigure}
\end{minipage}                        
\label{fig:charac}
\caption{Characterization of Microservices for a fixed input size from Djinn\&Tonic Benchmark Suite.}
\end{figure}
Figure~\ref{fig:slack} shows the breakdown of execution times per stage for 4 different microservice-chains. A detailed description of all the microservices used within the applications are  given in Table~\ref{tbl:microservices}. Consider the Detect Fatigue application shown in Figure~\ref{fig:slack}, 
It can be seen that 81\% of the total execution time is dominated by stage-1, whereas the other staged together take less than 20\% of the total time. A similar trend of non-uniform execution times is observed for the other three applications as well. Hence, it would be ideal to employ per-stage intelligent batching rather than uniformly batching requests across all stages, because the latter would lead to poor container utilization. 

To effectively exploit this per-stage variability described above, there are two assumptions: (i) the execution times for each stage of an application has to be known apriori, and (ii) the execution times should be predictable and not have large variations. The first assumption can be held true for serverless platforms because the applications are hosted as functions prior to the execution. A simple offline profiling can estimate the execution times of the functions. The second assumption also holds especially for ML-based applications because the ML-models use fixed-sized feature vectors, and exhibit input-independent control flow~\cite{swayam}. Therefore, the major sources for execution time variability come from application-independent delays that are induced by (i) scheduling policy or (ii) interference due to request co-location on the same servers. To support this claim, we conduct a characterization of 8 ML-based microservices from Djinn\&Tonic suite~\cite{djinn}. As shown in Figure~\ref{fig:exec}, the standard deviation in execution time measured across 100 consecutive runs of each microservice is within 20ms. In this experiment, the input size (image pixels or speech query) for all the microservices are fixed. Note that execution will vary depending on the input size to each microservice (for eg, 256x256 vs 64x64 image for IMC application). In our experiments we find a linear relationship between the execution time and the input size for these applications.\\
\textbf{Why does slack arise?} 
Though user-facing applications can have varied runtimes, the SLO requirement is deterministic because it is based on human perceivable latency. Because these applications are typically short-running, considerable amount of slack will exist. If we know the end-to-end runtime, we can estimate slack as the difference between runtime and response latency. For example, consider the execution times of the four ML based applications shown in Figure~\ref{fig:slack}: (i) Detect Fatigue, (ii) Intelligent Personal Assistant (IPA), (iii) Image Recognition (IMG), and (iv) Face Security. It can be seen that the maximum execution time among them is well within 220ms. If the end-to-end response latency is fixed at 1000ms, it is evident that all applications have ample amount of slack, which can be used to batch requests together.\\ 
\textbf{Key takeaway:} \textit{
RM frameworks should capitalize on both --- variability of execution time across stages, as well as overall application slack, by determining the optimal batch size to queue requests at every stage. This, in turn, can lead to better bin-packing of requests into fewer containers improving overall container utilization.}
\section{Preamble to Fifer}
\label{sec:modeling}
This section introduces how a RM framework can benefit from addressing the above shortcomings. Our baseline is representative of a RM used in existing serverless platforms like AWS lambda~\cite{peeking}, which spawns new containers for every request if there are no idle containers (as explained in Section~\ref{sec:coldstart}). On top of these RM frameworks, one can additionally batch the requests by queuing them at every stage of an application, which we name as Request Batching RM (RBRM). Contrary to existing RMs~\cite{azure,fission}, in RBRM, instead of statically assuming the batch size, we calculate the batch size (\texttt{B\_size}) as a function of execution time (\texttt{Stage\_Exec\_Time}) and available slack for each stage (\texttt{Stage\_Slack}), as shown in equation~\ref{eq:1} below.        
\begin{equation}                      
\small
{B\_size} \Leftarrow {Stage\_Slack} / {Stage\_Exec\_Time}
\label{eq:1}
\end{equation}
Based on \texttt{B\_size}, we can queue different number of requests at each stage. 
\begin{figure}
\begin{minipage}{0.99\linewidth}
\centering
\includegraphics[width=0.95\textwidth]{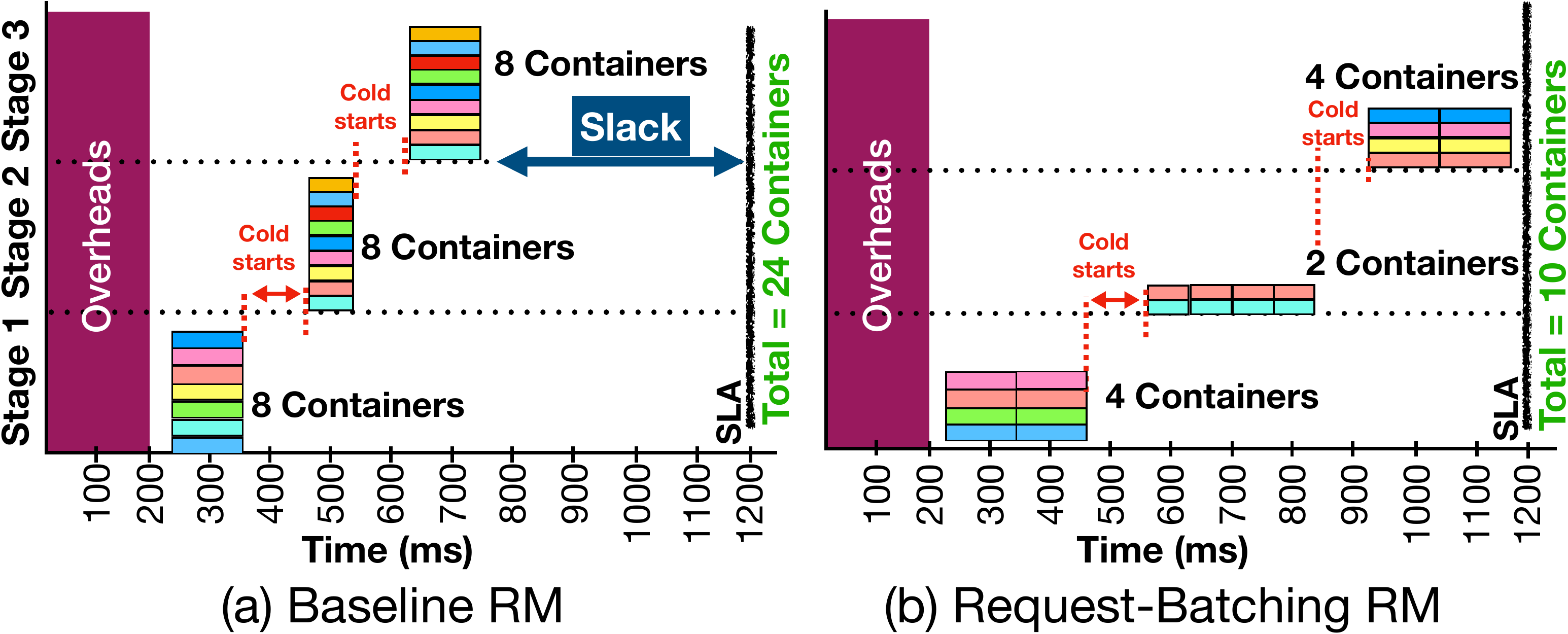}
\end{minipage}
\caption{An example representation showing the working of Baseline vs Stage-Aware batching enabled RM framework.}
\label{fig:architecture}
\end{figure}
Figure~\ref{fig:architecture} shows an example of how the baseline RM compares against RBRM for incoming requests. It can be seen that while the baseline (Figure~\ref{fig:architecture}a) spawns a total of 24 containers (with 8 containers per stage), the RBRM exploits slack by consolidating requests and spawns only 10 containers in total. Note that RBRM does not violate SLOs despite batching the requests. In Figure~\ref{fig:architecture}b, all requests are batched into 10 containers and the number of requests batched per container at every stage depends on the execution time and available slack at each stage~\footnote{We make a reasonable assumption here that, the tenants can provide their SLO requirements to serverless providers, thus enabling the provider to estimate slack.}.
Additional containers would be spawned if the arrival rate increases over time. 

It is important to note that queuing and batching still cannot help in hiding the cold-start latencies encountered (shown in Figure~\ref{fig:architecture}) when spawning new containers. These are exacerbated especially when there is dynamism in request rate. While cold-start latencies can be reduced by OS-level optimizations~\cite{234835}, the only way to hide them entirely is by proactively spawning containers. Balancing the aggressiveness of proactively spawning new containers and queuing requests at existing ones is crucial for achieving high container utilization and low SLOs, hence this act hinges solely on the prediction model adopted.\\
\textbf{Key takeaway:} \textit{Request batching can minimize containers spawned while avoiding SLO violations, but cannot hide cold-starts. SLO violations caused by cold starts can be avoided by provisioning containers in advance by predicting the future load, but reaping the benefits is contingent upon the accuracy of the prediction model used.}

\section{Overall Design of Fifer} 
\label{sec:scheme}
\begin{figure}
\begin{minipage}[t]{0.99\linewidth}
\begin{subfigure}[t]{0.99\linewidth}
\centering
\includegraphics[width=0.84\textwidth]{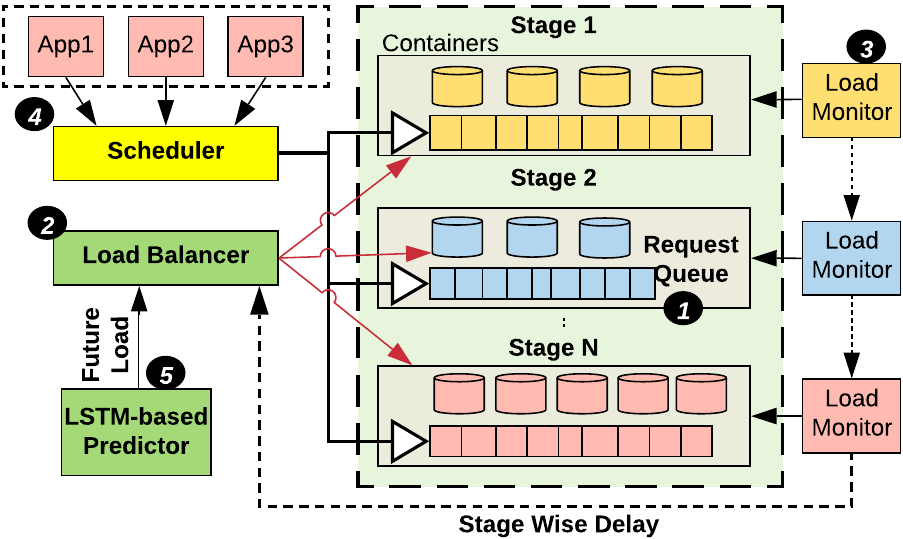}
\end{subfigure}
\end{minipage}
\caption{An overview of \textit{Fifer}.}
\label{fig:fifer}
\end{figure}

Motivated by our observations, we design a novel RM framework \textit{Fifer} to manage function-chains on serverless platforms efficiently. 
Figure~\ref{fig:fifer} depicts the high-level design of \textit{Fifer}. The key components of the design are explained in detail in the below subsections using circled annotations.
\subsection{Estimating execution time and slack}
\label{sec:estimation}
As briefly mentioned in Section~\ref{sec:modeling}, by knowing the available slack and execution time at each stage, we can accurately determine the number of requests that can be executed in a batch in one container. We conduct offline profiling to calculate the runtimes of all microservices used in six commonly used ML-based applications from the \emph{Djinn\&Tonic}~\cite{djinn} benchmark suite (briefly explained in Section~\ref{sec:exec}). Using the offline values, we build an estimation model using linear regression, which accurately generates a Mean Execution Time (MET) of each service for a given input size (shown in Table~\ref{tbl:microservices}). We do not use larger inputs which violate our SLO requirements. This model is added as an offline component to \emph{Fifer}. \\
\textbf{Slack Distribution:} To accurately estimate the slack for every application, we fix the SLO (response latency) as 1000ms, which is the maximum of 5 $\times$ execution\_time~\cite{swayam} of all the applications used in our workloads. By knowing the overall application execution time and response latency, the difference is calculated as \textbf{slack} for the application. To determine the slack for every stage of the application, we distribute the total slack to individual stages. This can be done in two ways, (i) the overall slack can be equally divided (ED) and allocated to each stage or (ii) the overall slack can be proportionately allocated with respect to the ratio of the execution time of each stage. In \emph{Fifer}, we use proportionate slack allocation for each stage, as it is known to give better per-stage utilization compared to ED~\cite{grandslam}.
\subsection{\textbf{Load Balancing}}
\label{sec:load-balancer}
\emph{Fifer} makes use of a request queue, which holds all the incoming tasks for each stage~\tcircled{1}. We design a load balancer (LB)~\tcircled{2} along with a load monitor that are integrated to each stage (LM)~\tcircled{3} for efficiently scaling containers for the application. Since we know the execution time and available slack, the LB can calculate the batch size (B\_size) for each stage using Equation~\ref{eq:1}. \\
\textbf{Dynamic Reactive Scaling Policy}:
To accurately determine the number of containers needed at every stage which is a function of B\_size and queue length, we need to periodically measure the queuing delay due to batching of requests. As shown in Algorithm~\ref{algo}\circled{a}, for a given monitoring interval at every stage, the LM monitors the scheduled requests in the last 10s to determine if there are any SLO violations due queuing delays. This is because there are not enough containers to handle all the queued requests. In that case, we estimate the additional containers needed using the \emph{Estimate\_Containers} function. By knowing the B\_size and number pending requests in the Queue (PQ$_{len}$), the function can estimate the number of containers N$_c$ = {PQ$_{len}$} / {B\_size}.

However, based on PQ$_len$, spawning a new container would be deemed unnecessary, if the time taken to service the requests by queuing on existing containers is lower than the cold-start delays. Therefore, the function takes into account, the delay incurred in terms of queuing the request for a longer time  vs cold start (C$_d$). The queuing delay threshold D$_f$ for Stage S is calculated using total number of requests that can be executed without violating SLOs (L), and total time required to satisfy all pending requests (T$_d$) as shown below:
\begin{align*}
L &= \sum_{i=1}^{N}{Bsize_i}, & T_d &= PQ_len \times S_r, & D_f &= \frac{T_d}{L} 
\end{align*}
where S$_{r}$ is the per stage response latency, N is the number of containers in S.   S$_{r}$ for a stage is defined as the sum of its allocated slack and execution time. If D$_f$ > C$_d$, then LB spawns additional containers (N$_c$) for each stage. We refer to this as dynamic reactive scaling (\emph{RScale}) policy.\\
\textbf{Stage-aware Container Scaleout}:
Since each stage of an application has asymmetrical running times (as shown in Figure~\ref{fig:slack}), the number of containers needed at every stage would be different. The baseline RM is not aware of this disparity. However, \emph{Fifer} is inherently stage-aware because it employs a proportionate slack allocation policy. This, in turn, results in having similar batch sizes for the containers at every stage though they have disproportionate execution times. Furthermore, the LM in \emph{Fifer} estimates the queuing delay for every individual stage by continuously monitoring the load. This, in turn, aids in better stage-wise container scaling as opposed to uniformly scaling containers. 
\subsection{Function Scheduling}
Apart from dynamically scaling the number of containers needed to host the requests per stage, we also need to design a scheduling policy to select the appropriate request from the queue of each stage. One important concern here is that a single application developer can host multiple types of applications from which some might share common functions (stages)~\footnote{It is be noted that serverless platforms do not share microservices across tenants. Doing so would violate the security and isolation guarantees.}. In such cases, the request queue for shared stages will have queries from different applications where the available slack for each application will be different depending on the overall execution time of the application. Therefore, executing the requests in FIFO order will lead to SLO violations. To ensure SLOs of shared functions, we employ a  Least Slack First (LSF) scheduling policy (shown in Algorithm \ref{algo}\circled{b}). \textit{Fifer} makes use of LSF such that it executes the application query with the least available slack from the queue at every stage. LSF helps to prioritize requests which have less slack and, at the same time, avoids starvation of requests in the queue. 
Since sharing microservices is not our primary focus in this work, we do not discuss the trade-offs involved in using other sharing specific scheduling policies.

\subsection{Bin-Packing to increase Utilization} 
\subsubsection{Greedy Container Scheduling} In order to increase utilization, we need to ensure minimal number of idle containers at any given point, which depends on the scheduling policy~\tcircled{4}. In \emph{Fifer}, we design a scheduling policy such that, each stage will submit the request to the container with the least remaining free-slots where the number of free-slots is calculated using the container's batch-size. 
In addition, we use a timeout period of 10 minutes to remove idle-containers which have not serviced any requests for that period. Hence, employing a greedy approach of choosing containers with the  least-remaining free-slots (shown in Algorithm~\ref{algo}\circled{d}) as a scheduling candidate will in turn result in early scale-down of lightly loaded containers.
\subsubsection{Greedy Node Selection}
The containers used to host functions are themselves hosted on servers, which could be VMs or bare-metal servers. In \emph{Fifer}, similar to the function scheduling policy, we greedily schedule containers on the least-resource-available server. The servers are numbered from 1 to n where n is the number of available servers. We tune the cluster scheduler to assign containers to the lowest numbered server with the least available cores that can satisfy the CPU and memory requirement of the container. As a result, the unused cores will only be consuming idle power, and also the servers with all cores being idle can be turned after some duration of inactivity. Consequently, this can translate to potential savings in cluster energy consumption as a result of bin-packing all active containers on to fewer servers.
\begin{algorithm}[t]
\caption{Stage\_Aware + Slack\_Aware + Prediction}
\begin{algorithmic}[1]
\Procedure{Dynamic\_Reactive\_Scaling(\textit{Stages})}{}~\circled{a}
\For{\textit{stage in $\forall Stages$}}
\State$\textit{delay} \gets \textit{\textbf{Calculate\_Delay}({last\_10s\_jobs})}$

\If{\textit{delay $\ge$ stage.slack}} 
\State$\textit{est\_containers} \gets \textit{\textbf{Estimate\_Container}()}$
\State$\textit{stage.containers.append}(\textit{est\_containers})$
\EndIf
\EndFor
\EndProcedure 
\hrulefill
\Procedure{Predictive\_Stage\_Aware(\textit{Stages})}{}~\circled{e}

\State$\textit{load} \gets Measure\_Load(last\_100\_jobs)$
\For{\textit{stage in $\forall Stages$}}
\State$\textit{batchSize} \gets \textit{stage.batchSize}$
\State$\textit{current\_req} \gets \textit{len(stage.containers) $*$ batchSize}$
\State$\textit{Fcast} \gets \textit{\textbf{LSTM\_Predict}(\textit{load})}$
\If{\textit{Forecast $\ge$ current\_requests}}
\State$\textit{est\_containers} \gets \textit{(Fcast - current\_req)}$
\State$\textit{est\_containers} \gets \textit{est\_containers / batchSize)}$

\State$\textit{stage.containers.append}(\textit{containers\_needed})$
\EndIf
\EndFor
\EndProcedure 
\end{algorithmic}
\label{algo}
\end{algorithm}

\subsection{Proactive Scaling Policy}
\label{sec:prediction}
It is to be noted that, the queuing delay estimations and scaling based on runtime delay calculations would still lead to sub-optimal container spawning, especially if the future arrival rate is not known.  
Hence, in \textit{Fifer}, we use a load prediction model~\tcircled{5},  which can accurately forecast the anticipated load for a given time  interval. Using the predicted load, \textit{Fifer} proactively spawns new containers at every stage. 

As shown in Algorithm~\ref{algo}\circled{e}, for every monitoring interval, we forecast the estimated number of requests based on past arrival rates. For each stage, if the current number of containers available is not sufficient to handle the forecast request load, \textit{Fifer} proactively spawns additional containers. This proactive scaling policy is complementary to the dynamic reactive policy at each stage. If the prediction model can accurately predict the future load, then it would not result in SLO violations as the necessary number of containers would be spawned in advance. However, in the case of mispredictions, the reactive policy would detect delays at the respective stages and spawn additional containers with cold-starts.
We next explain in detail the prediction model used in \emph{Fifer}.

\quad To effectively capture the temporal nature of request arrival scenario in date-centers, we make use of a Long Short Term Memory Recurrent Neural Network (LSTM) model~\cite{lstm}. LSTMs are known to provide a state-of-the-art performance for many popular application domains, including Stock Markets forecast and language processing. For a periodic monitoring interval (T) of 10s, \textit{Fifer} samples the arrival rate in adjacent windows of size W\textsubscript{s} (5s) over the past 100 seconds. It keeps track of the maximum arrival rate at each window and calculates the global maximum arrival rate. Using this global arrival rate, the model predicts the number of containers as a maximum in the future window of size W\textsubscript{p}.  The interval (T) is set to 10 seconds since average container start-up latency ranges between 1s and 10s. The prediction window (W\textsubscript{p}) is set to 10 minutes since 10 minutes of future trend is sufficient to expose the long term trade-offs.  Short-term load fluctuations would still be captured within the 10s interval.
\begin{figure}
\begin{minipage}[t]{0.99\linewidth}
\centering
\begin{subfigure}{0.49\linewidth}
 \includegraphics[width=0.99\textwidth]{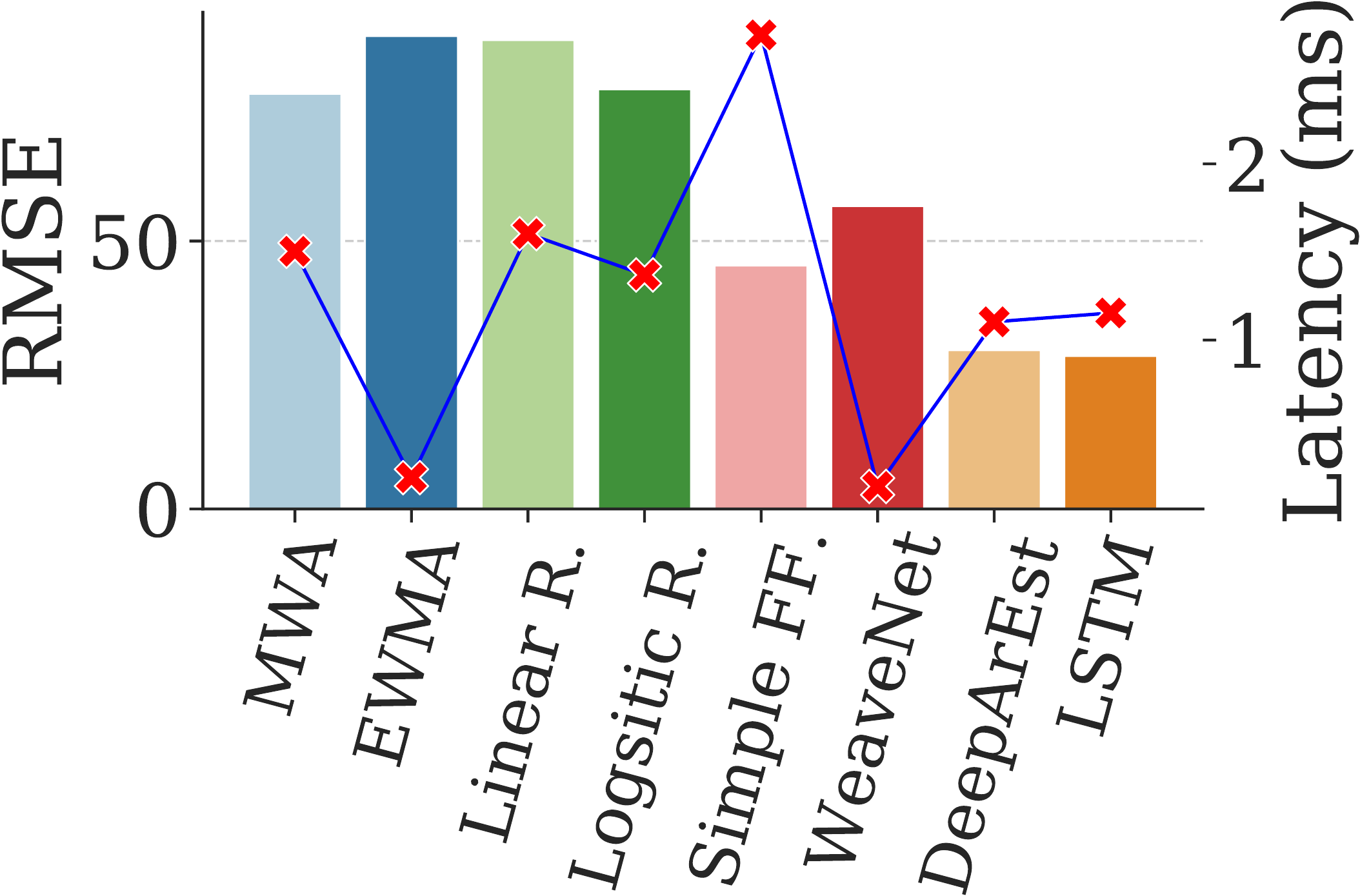}
 \caption{RMSE and Latency(ms).}
\label{fig:ml-charac}
\end{subfigure}
\begin{subfigure}{0.49\linewidth}
 \includegraphics[height=2.6cm,width=0.99\textwidth]{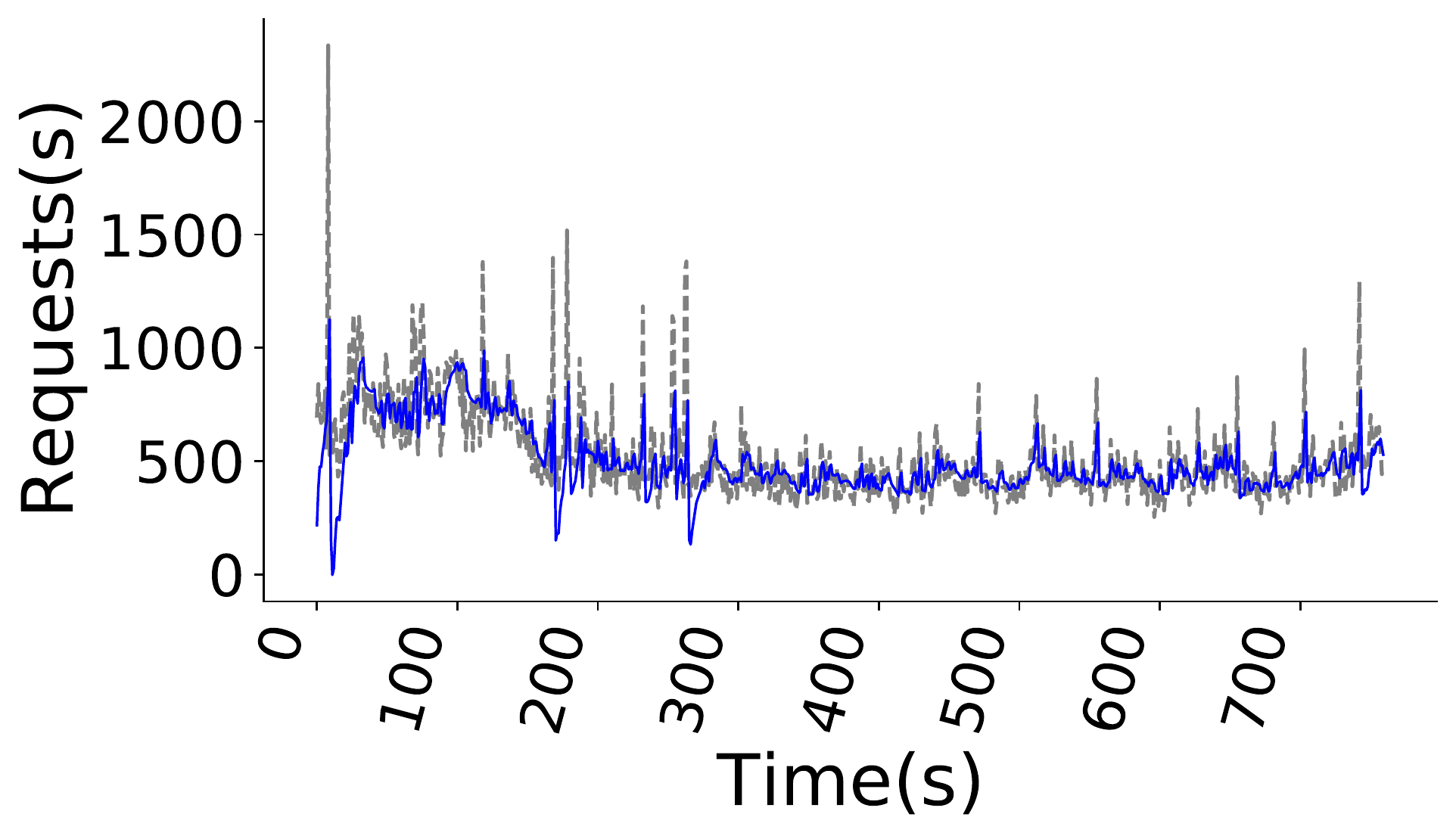}
\caption{Prediction accuracy of LSTM model.}
\label{fig:lstm}
\end{subfigure}
\end{minipage}    
\caption{Comparing different prediction models.}
\end{figure} 
\subsubsection{\textbf{Prediction Model Design.}}
We also quantitatively justify the choice of using LSTM by a doing a brick-by-brick comparison of the trade-offs of using different non-ML based and ML-based models on a given input arrival trace. We use four non-ML models, namely Moving Window Average (MWA), Exponential Weight Moving Average(EWMA), Linear Regression (Linear R.), and Logistic Regression (Logisitic R.). These models are continuously fitted over requests in last t-100 seconds for every T. In addition, we use four ML models (Simple Feed Forward Network, WeaveNet, DeepAREstimator and LSTM) that are pre-trained with 60\% of the WITS~\cite{wits} arrival trace input as the training set. We employ a time-step based prediction on the ML models as described in the above sub-section. Figure~\ref{fig:ml-charac} plots the Root Mean Squared Error (RMSE) and latency incurred by 8 different prediction models.  It can be seen that LSTM has the least RMSE values. To verify the same, we plot the accuracy of the LSTM model for WITS trace (Figure~\ref{fig:lstm}). It is evident that the model predicts requests accurately (85\% from our experiments) for the given test set duration of 800s.
\section{Implementation and Evaluation}
\label{sec:implementation}
We have implemented a prototype of the proposed \emph{Fifer} framework using open-source tools for evaluating the design with synthetic and real-world traces. The details of the implementation are described below. 

\subsection{{\emph{Fifer} Prototype Implementation}}
\emph{Fifer} is implemented on top of \emph{Brigade}~\cite{brigade} using 5KLOC of JavaScript. \emph{Brigade} is an open-sourced serverless workflow management framework developed by Microsoft Azure~\cite{brigade-azure}. \emph{Brigade} can be deployed on top of a Kubernetes~\cite{kubernetes} cluster that handles the underlying mechanisms of pod(container) creation, orchestration, and scheduling. \emph{Brigade}, by default, creates a worker pod for each job, which in turn handles container creation and scheduling of tasks  within the job and destroys the containers after job completion. Henceforth, we refer to a function chain as a ``job'' and stages within the job as ``tasks''. To cater to \emph{Fifer}'s design, we modify \emph{Brigade} workers to persist the containers for every task after job completion such that they can be reused for scheduling tasks of other jobs. We implement a global request queue for every stage within the job which holds all the incoming tasks before being scheduled to a container in that stage. Each container has a local queue of length equal to the number of free-slots in the container. 

We integrate a \emph{mongodb}~\cite{mongodb} database to maintain job-specific statistics (creationTime, completionTime, scheduleTime,  etc) and container-specific metrics(lastUsedTime, batch size, etc), which can be periodically {\em queried} by the worker pod and load-balancer. As an offline step, for every function chain the following are added to the database, (a) the response latency, (b) the sequence of stages, (c) estimated execution time, and (d) the slack per stage (calculated as described in Section~\ref{sec:estimation}). Using these values, each container of a stage can then determine its free-slots. \\ 
\noindent{\textbf{Pod Container Selection:}}  At runtime, the worker pod queries the database to pick a pod with the least number of free slots to schedule the task. Once a pod is selected, the task is added to its local-queue, and the free-slots of the pod are updated in the database. The same process is applied to every subsequent task of the job. \\
\noindent{\textbf{Load Balancer:}} We designed a \emph{Python} daemon (1K LOC),  which consists of a load monitor (LM) and a load predicto (LP). The LM periodically measures the queuing delay at the global queue of each stage and  spawns additional containers if necessary (described in Section~\ref{sec:load-balancer}). The LP predicts the request-load using the LSTM model. The model was trained using Keras~\cite{keras} and Tensorflow~\cite{tensorflow}, over 100 epochs with 2 layers, 32 neurons, and batch size 1. The daemon queries the job\_arrival information from the database, which is given as input to the prediction model. Recall that the details of the prediction were described in Section \ref{sec:prediction}.\\
\noindent{\textbf{Node/Server Selection:}} In order to efficiently bin-pack containers into fewer nodes, we make modifications to the \emph{MostRequestedPriority} scheduling policy in Kubernetes such that it always chooses the node with the least-available-resources to satisfy the Pod requirements. For our experiments, each container requires 0.5 CPU-core and memory within 1GB. Hence, we set the CPU limit for all containers to be 0.5. We determine idle cores in a node by calculating the difference between number of cores in a node and the sum of cpu-shares for all allocated pods in that node.

\subsection{{Large Scale Simulation}} 
To evaluate benefits of \emph{Fifer} for large scale systems, we built a high-fidelity event-driven simulator using container cold-start latencies, loading times of container images and function transition times from our real-system counterpart. Using synthetic traces in both the simulator and the real-system, we verified the correctness of the simulator by comparing and correlating various metrics of interest.  
\begin{table}
\begin{minipage}{.45\linewidth}
\begin{center}
\footnotesize
\resizebox{\textwidth}{!}{
 \begin{tabular}{||p{1.65cm}| p{1.65cm}||} 
 \hline
  \textbf{Hardware}  & \textbf{Configuration}\\
 \hline
 \textit{CPU}  & Xeon R Gold-6242 \\
 \hline
 \textit{Sockets}  & 2 \\
 \hline
 \textit{Cores($\times$)threads} & 16 $\times$ 2 \\  
  \hline
  \textit{Clock} & 2.8 Ghz \\
  \hline
  \textit{DRAM} & 192 GB \\
  \hline

  \end{tabular}}
\end{center}

\caption{Hardware config.}
\label{tbl:hw-config}

\end{minipage}%
 \hspace{2mm}
\begin{minipage}{.45\linewidth}
\begin{center}
\footnotesize
 \resizebox{\textwidth}{!}{%
 \begin{tabular}{|p{1.8cm}| p{1.6cm} ||} 
 \hline
   \textbf{Software} & \textbf{Version} \\
    \hline
    \textit{Ubuntu} &  16.04 \\
   \hline
   \textit{Kubernetes} & 1.18.3 \\ 
   \hline
   \textit{Docker} &  19.04 \\
    \hline
   \textit{MongoDB} & 2.6.10  \\ 
    \hline
   \textit{Python} &  3.6 \\
  \hline
  \textit{Tensorflow} &  2.0 \\
  \hline
\end{tabular}}
\end{center}

\caption{Software config.}
\label{tbl:sw-config}
\end{minipage}%
\end{table}
\begin{figure}
\begin{minipage}{0.99\linewidth}
\begin{subfigure}{0.45\linewidth}
    \centering
    \includegraphics[width=1\textwidth]{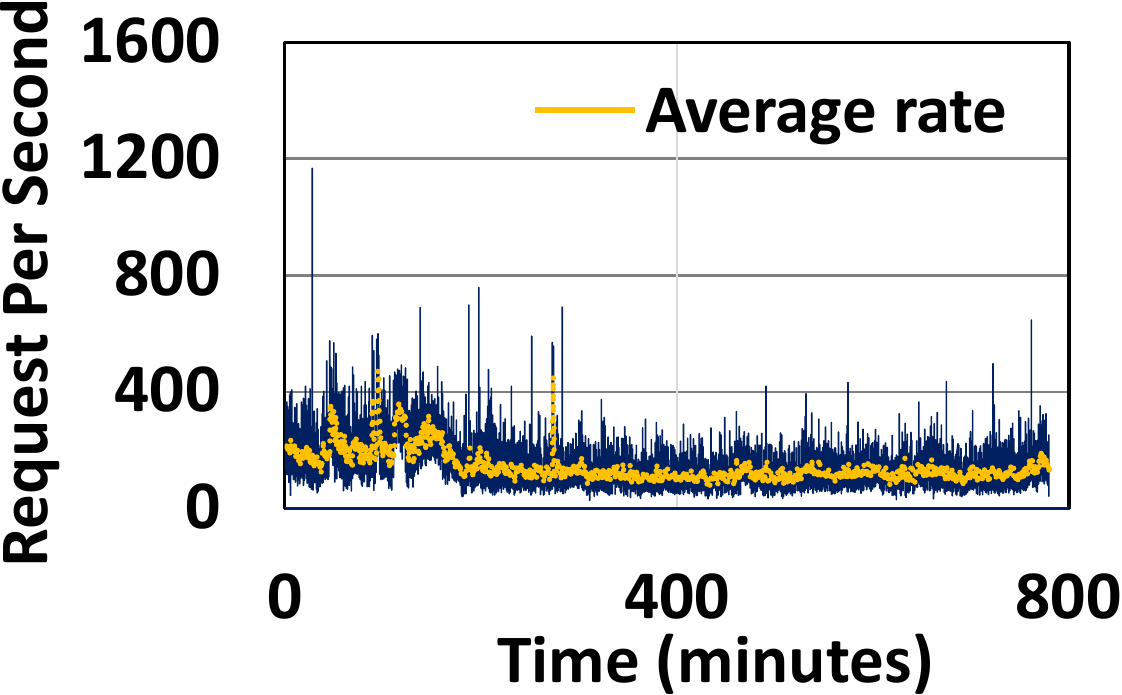}
    \caption{WITS Trace.}
        \label{fig:wits-trace}
    \end{subfigure}
\begin{subfigure}{.45\textwidth}
    \centering
    \includegraphics[width=1\textwidth]{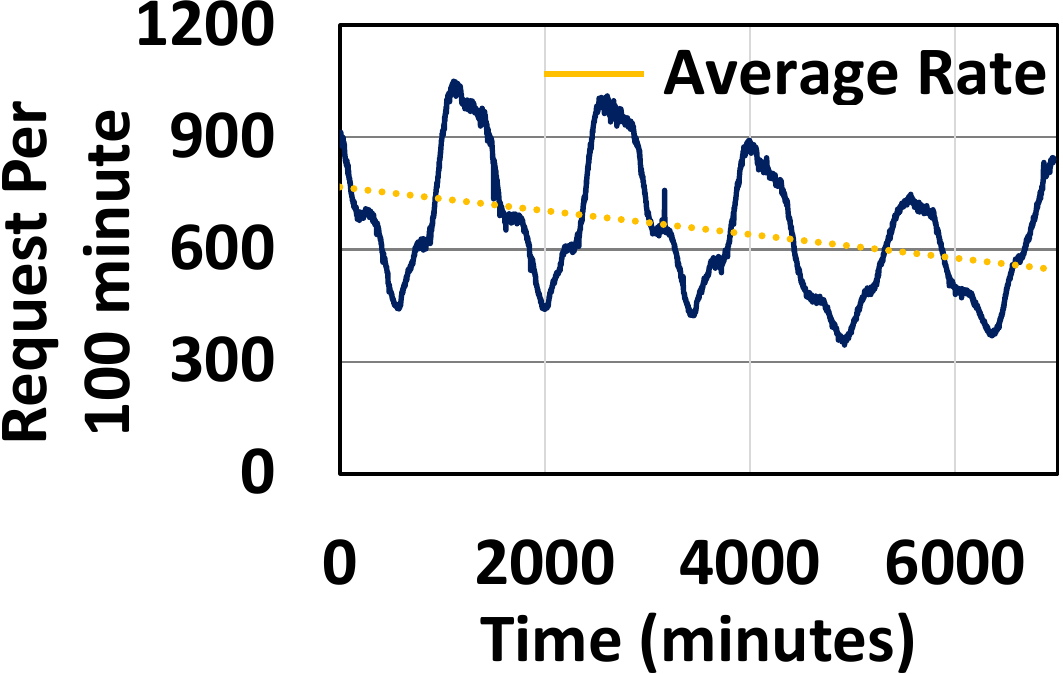}
    \caption{Wiki Trace.}
    \label{fig:wiki-trace}
\end{subfigure}
\end{minipage}
\vspace{-2mm}
\caption{Job Request Arrival Traces.}
\label{fig:trace}
\end{figure}
\begin{table}
\resizebox{\linewidth}{!}{%
\begin{tabular}{||l|l|l|l||}
\hline
Domain &
  ML application &
  ML Model &
  \begin{tabular}[c]{@{}l@{}}Avg. Exec \\ Time (ms)\end{tabular} \\ \hline
\multirow{5}{*}{Images Services} &
  Image Classification (IMC) &
  Alexnet &
  43.5 \\ \cline{2-4} 
 & Human Activity Pose (AP)       & DeepPose & 30.3  \\ \cline{2-4} 
 & Human Segmentation (HS)        & VGG16    & 151.2 \\ \cline{2-4} 
 & Facial Recognition (FACER)     & VGGNET   & 5.5   \\ \cline{2-4} 
 & Face Detection (FACED)         & Xception & 6.1   \\ \hline
Speech Services &
  \begin{tabular}[c]{@{}l@{}}Auto Speech Recognition (ASR)\end{tabular} &
  NNet3 &
  46.1 \\ \hline
\multirow{3}{*}{\begin{tabular}[c]{@{}l@{}}Natural\\Language\\ Processing\end{tabular}} &
  Parts of Speech Tagging (POS) &
  SENNA &
  0.100 \\ \cline{2-4} 
 & Name Entity Recognition (NER) & SENNA    & 0.09  \\ \cline{2-4} 
 & Question Answering (QA)        & seq2seq  & 56.1  \\ \hline
\end{tabular}}
\caption{Description of Microservices (Functions) used in \emph{Fifer}.}
\label{tbl:microservices}
\end{table}
\subsection{Evaluation Methodology} 
\label{sec:evaluation}
We evaluate our prototype implementation on an 80 compute core Kubernetes cluster. We use one dedicated node as the head node. Each node is a Dell PowerEdge R740 server with Intel CascadeLake Xeon CPU host. The details of the single node hardware and software configuration are listed in Table~\ref{tbl:hw-config} and \ref{tbl:sw-config}. We use \textit{Kubernetes} as the resource orchestrator. The \emph{mongodb} database~\cite{mongodb} and \emph{Python} daemon reside on the head node. For energy measurements, we use an open-source version of Intel Power Gadget~\cite{power} measuring the energy consumed by all sockets in a node.\\
\textbf{{Load Generator:}} \label{sec:loadgen}We use  different traces which are given as input to the load generator. Firstly, we use synthetic Poisson-based request arrival rate with average arrival $\lambda=50$. Secondly, we use real-world request arrival traces  from Wiki~\cite{wiki} and WITS~\cite{wits} (shown in Figure~\ref{fig:trace}). As shown in Figure \ref{fig:wits-trace}, the WITS trace has a large variation in peaks (average=300req/s, peak=1200 req/s) when compared to the Wiki trace. The wiki trace (average= 1500 req/s) exhibits the typical characteristics of ML inference workloads, containing recurring patterns (e.g., hour of the day, day of the week), whereas the WITS trace contains unpredictable load spikes (e.g., black-Friday shopping). \emph{Based on the peak request arrival rate, the simulation expands to match up to the capacity of a 2500 core cluster (30$\times$ our prototype cluster)}.  

Each request is modelled after a query, which could be one among the four   applications (microservice-chains), as shown in   Table~\ref{tbl:applications}. Each application is compiled as a workflow program in Brigade, which invokes each microservice container in a sequence. The applications consist of well-known microservices derived from the \textit{Djinn\&Tonic}~\cite{djinn} benchmark suite (see Table~\ref{tbl:microservices}). These include microservices from a diverse range of domains like image recognition, speech recognition, and language processing. All our  microservices utilize \textit{Kaldi~\cite{kaldi}, Keras~\cite{keras} and Tensorflow~\cite{tensorflow}} libraries.\\ \\
\begin{table}
\centering
\footnotesize
\resizebox{0.94\linewidth}{!}{
\begin{tabular}{||c|c|c||}
\hline
\textbf{Application Type} & \textbf{ 
Microservice-chain} & \textbf{Avg Slack(ms)} \\ \hline
Face Security & FACED $\Rightarrow$ FACER & 788 \\ \hline
IMG & IMC $\Rightarrow$ NLP $\Rightarrow$ QA & 700\\ \hline
IPA & ASR $\Rightarrow$
  NLP $\Rightarrow$ QA & 697 \\ \hline
Detect-Fatigue & HS $\Rightarrow$ AP $\Rightarrow$ FACED $\Rightarrow$ FACER & 572 \\ \hline
\end{tabular}}%
\caption{Microservice-Chains and their slack.}
\label{tbl:applications}
\end{table}
\textbf{{Workload}:}
We model three different workload mixes by using a combination of two applications as
shown in Table~\ref{tab:workloads}. 
\begin{wraptable}{r}{0.3\textwidth}
\footnotesize
\begin{tabular}{||c|c||}
\hline
\textbf{Workload} & \textbf{Query Mix} \\ \hline
Heavy & IPA, DETECT-Fatigue \\ \hline
Medium & IPA, IMG \\ \hline
Light & IMG, FACE-Security \\ \hline
\end{tabular}%
\caption{Workload Mix.}
\label{tab:workloads}
\end{wraptable} 
Based on the increasing order of total available slack for each  workload (avg. of both application's slack), we categorize them into \textit{Heavy}, \textit{Medium}, and \textit{Light}. Using the three workload-mix, we comprehensively analyze the scope of the benefits of \emph{Fifer} for different proportions of available slack. The individual slacks for every application are shown in Table~\ref{tbl:applications}.\\ 
\textbf{\normalsize{Container Configuration:}}
All the microservices shown in Table~\ref{tbl:microservices} are containerized as ``pods'' in Kubernetes. We set the imagePullPolicy for each pod such that the container image will be pulled from dockerhub by default when starting a new container. This captures the behaviour of serverless functions where function instances are loaded from external storage for every new container.\\ \\
\textbf{Metrics and Resource Management Policies:}
We evaluate our results by using the following metrics: (i) percentage of SLO violations, (ii) average number of containers spawned, (iii) median and tail latency of requests, (iv) container utilization, and (v) cluster-wide energy savings. The tail latency is measured as the $99^{th}$ percentile request completion times in the entire workload. We compare these metrics for \emph{Fifer} against \emph{Bline}, \emph{Sbatch} and \emph{BPred} resource-managers (RMs). In \emph{Sbatch}, we set the batch size by equal-slack-division policy and fix the number of containers based on the average arrival rates of the workload traces. \emph{BPred} is built on top \emph{Bline} along with the LSF scheduling policy and the EWMA prediction policy. Note that this is a faithful implementation of scheduling and prediction policy as used in Archipelago~\cite{archipelago}, which does not support request batching. Further, to effectively compare the combined benefits of the individual components of \emph{Fifer}, we do a brick-by-brick comparison of \emph{Fifer} (a) only with dynamic scaling policy (\emph{RScale}), and (b) combined with \emph{RScale} and proactive provisioning. Both these variants employ the LSF job scheduling policy, as well as the greedy container/node selection policy. It is also to be noted that \emph{Fifer} with \emph{RScale} policy is akin to the dynamic batching policy employed in GrandSLAm~\cite{grandslam}.  

\section{Results and Analysis}
\label{sec:results}
\subsection{Real-System Prototype}
We explain the results of our real-system implementation in this subsection.
\subsubsection{Minimizing Containers:}
\begin{figure}
\begin{minipage}[t]{0.99\linewidth}
\begin{subfigure}[t]{.49\textwidth}
\centering
\includegraphics[width=0.99\textwidth]{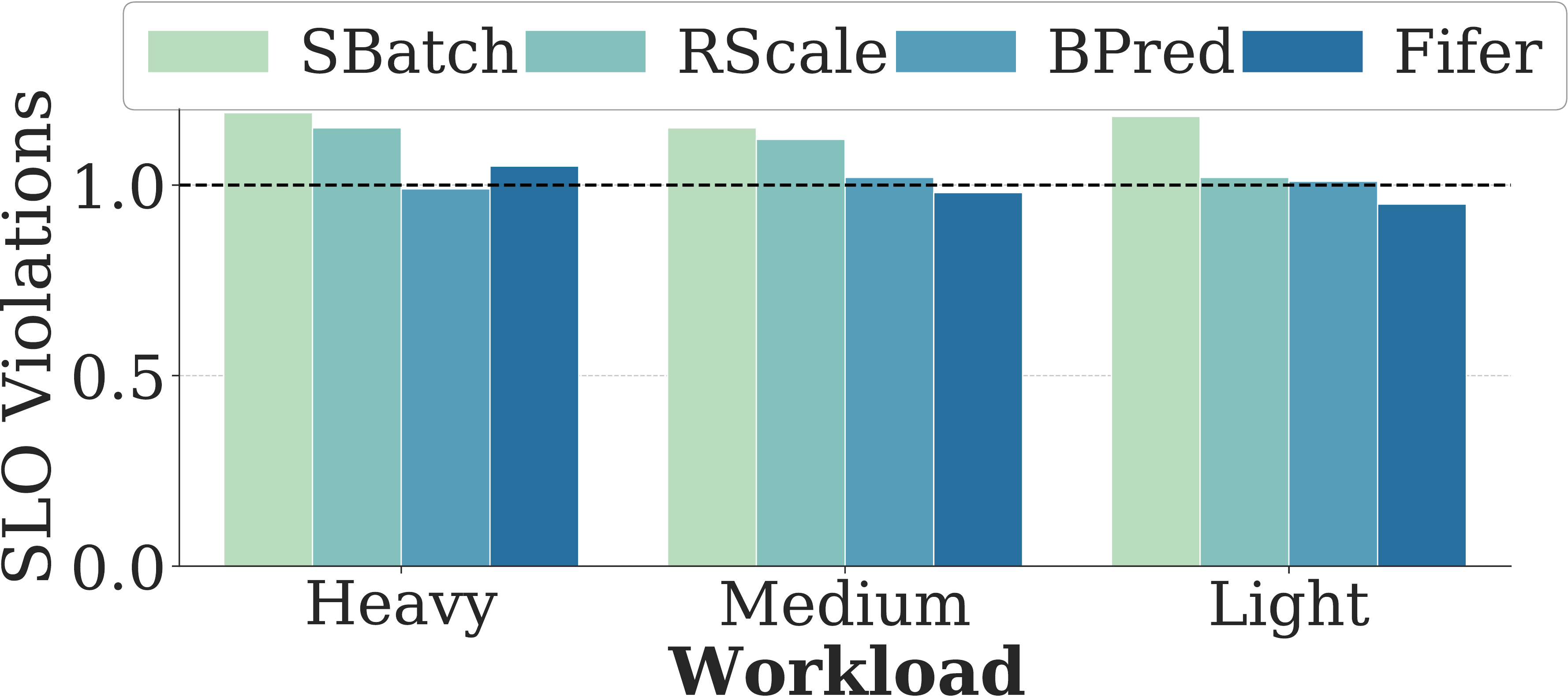}
\caption{SLO violations norm. to \texttt{Bline}.}
\label{fig:poisson-sla}
\end{subfigure}
\begin{subfigure}[t]{0.49\textwidth}
\centering
 \includegraphics[width=0.99\textwidth]{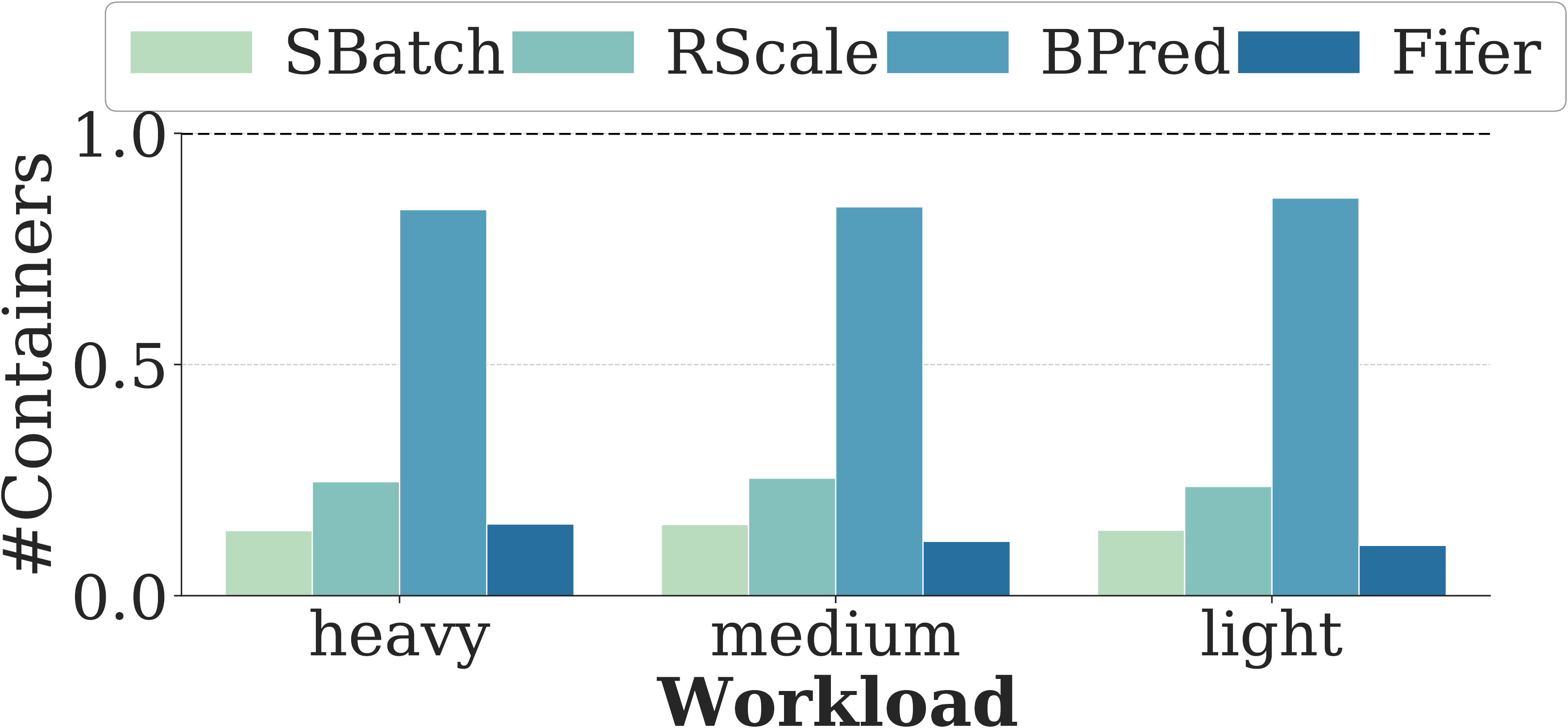}
\caption{Containers norm. to \texttt{Bline}.}
\label{fig:poisson-cont}
\end{subfigure}
\end{minipage}
\caption{\emph{Fifer} Prototype: Comparing SLO violations with number of containers spawned.\vspace{1mm}}
\end{figure}
Figure~\ref{fig:poisson-sla} and Figure~\ref{fig:poisson-cont} show the percentage of SLO violations and average number of containers spawned for the five different RMs across all three workload mixes. It is evident that \emph{Fifer} spawns the least number of containers on average compared to all other schemes except \emph{SBatch}. This is because \emph{SBatch} does not scale containers based on changes in request load. However, this results in  15\% more SLO violations for \emph{SBatch} when compared to Fifer. The \emph{Bline} and \emph{BPred} RMs inherently over-provision containers due to their non-batching nature, thus minimizing SLO violations. But the \emph{BPred} RM uses 20\% lesser containers on average when compared to \emph{Bline} due to proactive container provisioning. 
In contrast, both \emph{Fifer} and \emph{RScale} batch jobs to reduce the number of containers being spawned. While \emph{RScale} policy incurs 10\% more SLO violations than \emph{Bline} due to reactive-scaling when trying to minimize number of containers, Fifer does accurate proactive container provisioning thus avoiding SLO violations.
In short, \emph{Fifer} achieves the best of both worlds by combining benefits of batching (used by \emph{RScale}) and proactive scaling (used by \emph{BPred}).\\   
\subsubsection{Reduction in Latency:} 
\begin{wrapfigure}{l}{0.26\textwidth}
\captionsetup{justification=justified}
    \centering
    \includegraphics[width=.24\textwidth]{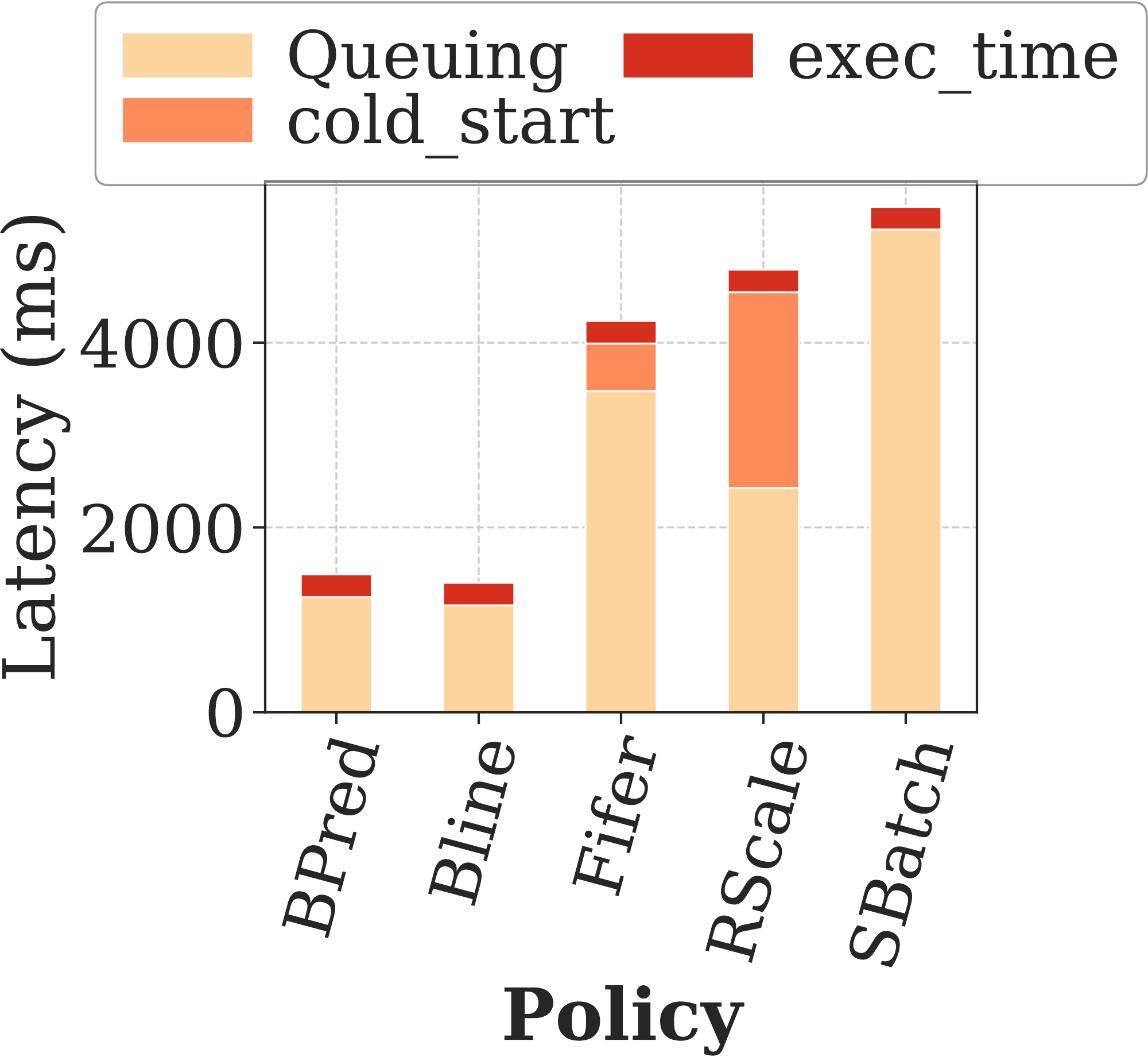}
        \caption{P99 Tail Latency.}
\label{fig:poisson-tail}
\end{wrapfigure}

\begin{figure}
\begin{minipage}[t]{\linewidth}
\centering
\begin{subfigure}[t]{.46\textwidth}
 \includegraphics[height=3.2cm,width=0.99\textwidth]{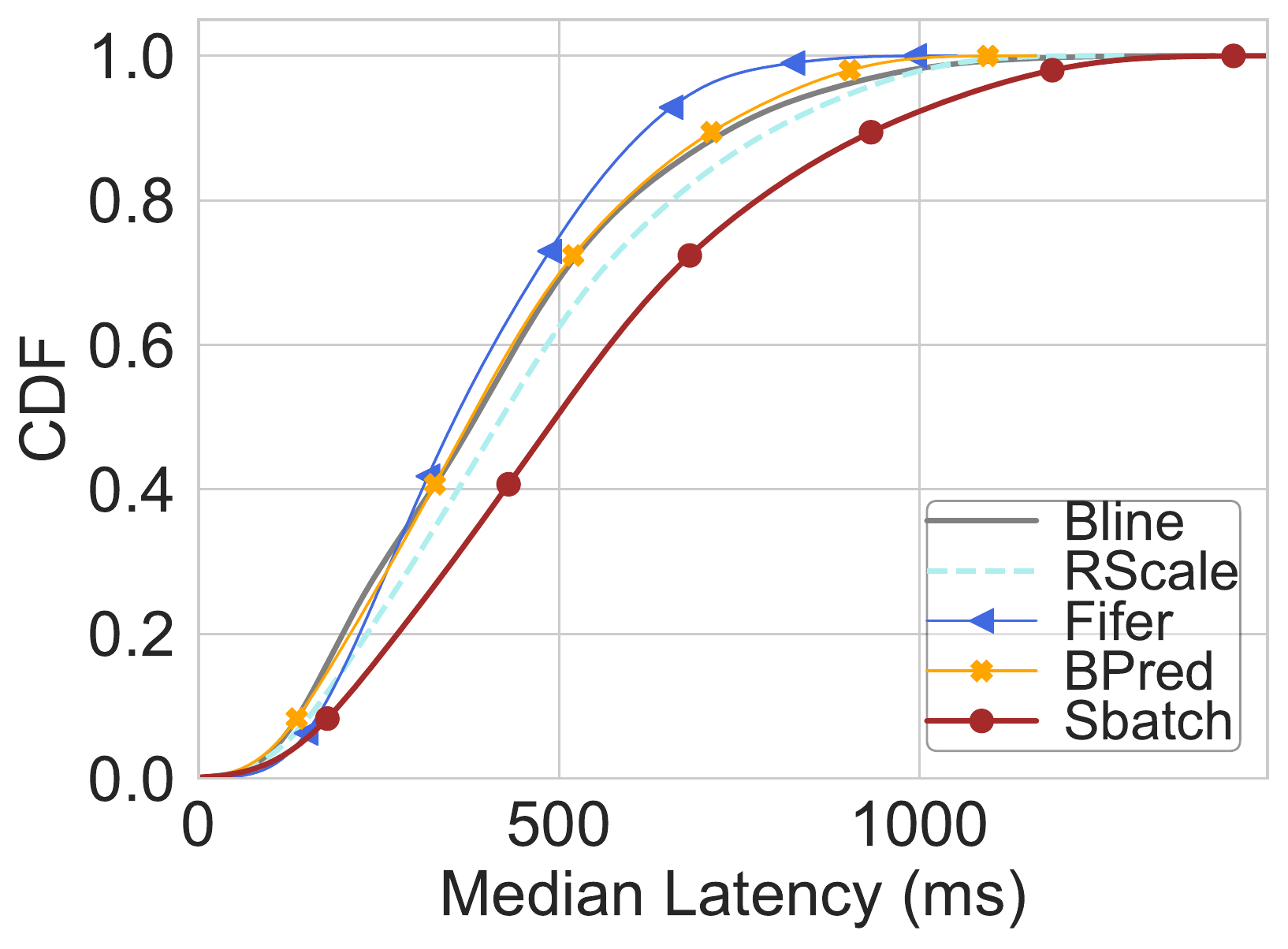}
\caption{Latency Distribution up to P95.}
\label{fig:poisson-median}
\end{subfigure}
\begin{subfigure}[t]{.45\textwidth}
 \includegraphics[height=3.2cm,width=0.99\textwidth]{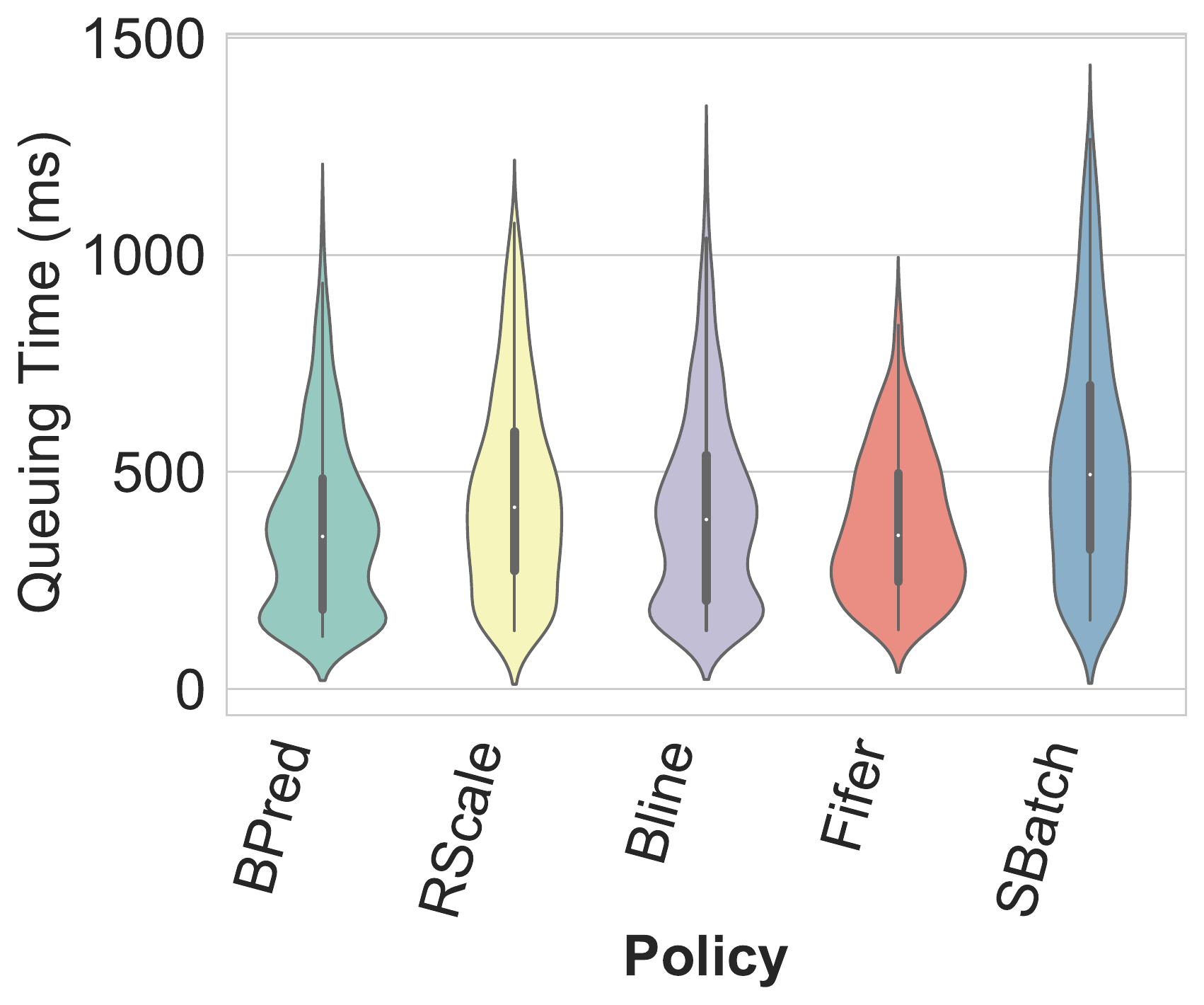}
\caption{Queuing time distribution.}
\label{fig:queue_time}
\end{subfigure}
\caption{Queuing time and response latency distribution for heavy workload-mix.}  
\end{minipage}                        
\end{figure}

Figure~\ref{fig:poisson-median} plots the CDF of total response latency up to P95 for heavy workload-mix. The breakdown of P99 tail latency is plotted separately in Figure~\ref{fig:poisson-tail}. 
We separate the response latency into execution time, cold-start induced delay, and batching induced delay. The batching induced delay is only for \emph{RScale} and Fifer policies. 
It can be seen that, the P99  is up to 3$\times$ higher for \emph{RScale} and \emph{SBatch} when compared to \emph{Bline} and \emph{Bpred}. This is because aggressive batching and reactive scaling do not handle load variations, which leads to congestion in the request queues. The \emph{Bpred} policy has lesser P99 compared to \emph{RScale} but in-turn it spawns 60\% more containers than \emph{RScale}. On the other hand, aggressive batching, along with proactive provisioning in \emph{Fifer} only lead to 2$\times$ higher P99 latency than both \emph{Bline} and \emph{Bpred}. Figure~\ref{fig:poisson-tail}. It can be seen that the delay due to cold-starts is much lower for \emph{Fifer} when compared to \emph{RScale}. 
This is because the number of reactively scaled containers are much lesser owing to the accurate estimation of future load by \emph{Fifer's} prediction model.

Since both the \emph{RScale} and \emph{Fifer}
RM enables batching of requests at each stage, the median latency of the requests is high compared to the \emph{\emph{Bline}} (shown in Figure~\ref{fig:poisson-median}). However, \emph{Fifer} utilizes the slack for requests at each stage, hence 99\% requests complete within the given SLO, despite having increased median latency. \vspace{-2mm}
\subsubsection{Breakdown of Key Improvements:}
\begin{wrapfigure}{l}{0.2\textwidth}
\captionsetup{justification=justified}
    \centering
    \includegraphics[width=.2\textwidth]{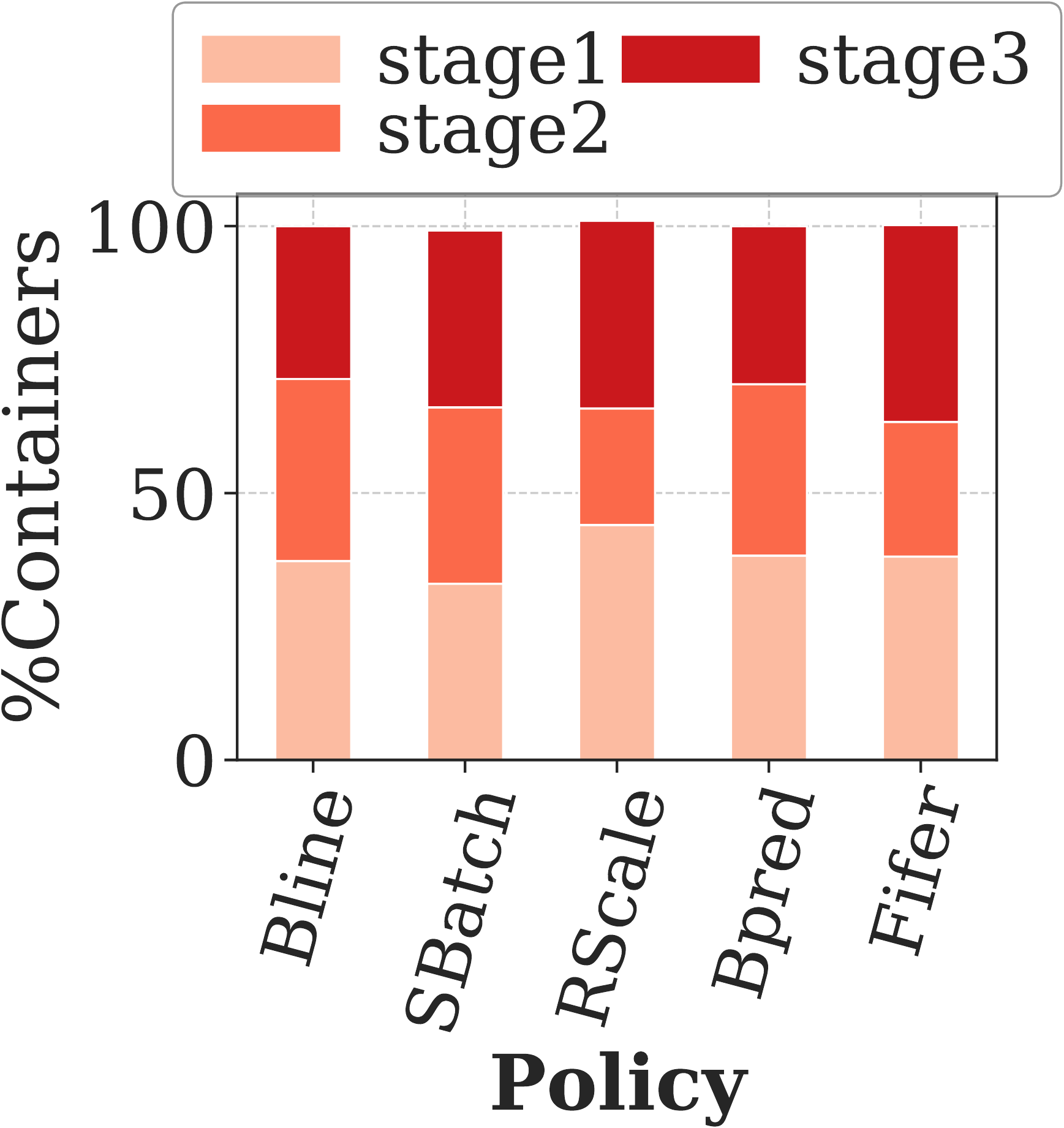}
        \caption{Distribution of Containers across stages of IPA application.}
\label{fig:stagewise}
\end{wrapfigure}

The major sources of improvements in \emph{Fifer} are (i) reduction of queuing delays due to proactive container provisioning and (ii) increased container utilization and better energy efficiency due to stage-aware container batching. We discuss these improvements in detail below. The stage-wise results are plotted for three stages of IPA application from heavy workload mix. The results are similar for other applications as well.\\ 
{\textbf{Effects of Queuing:}}
Figure~\ref{fig:queue_time} plots the queuing time distribution for heavy workload mix. It can be seen that the median queuing latency is very high for \emph{Fifer} (50-400ms), which indicates more requests are getting queued due to exploiting the slack of each stage. For the \emph{RScale} scheme, the median queuing latency is slightly higher than \emph{Fifer} (500ms)  because it leads to increased waiting times due to reactive spawning of containers with cold-starts. However, for both \emph{Bline} and \emph{BPred} RM, the latency distribution is irregular because the queuing latency will be higher or lower depending on the number of over-provisioned containers. \\
\noindent{\textbf{Stage-aware Batching and Scaleout}:} 
Figure~\ref{fig:stagewise} plots the stage-wise container distribution for all three stages. The execution time distribution for the stages was previously shown in Figure~\ref{fig:slack}. It can be seen that both  \emph{Bline} and \emph{BPred}  have more containers allocated for Stage-1 (ASR), which is the longest running stage (bottleneck) in the IPA application. However, the \emph{RScale} scheme spawns slightly higher containers for Stage-1 (44\%) and Stage-3 (35\%) when compared to Baseline. 
This is because  the proportionate slack allocation policy will evenly distribute the load across stages. Ideally, the distribution should be very close to \emph{Sbatch}, but reactive scaling of containers leads to many idle containers in each stage. \emph{Fifer}, on the other hand, spawns almost equal percentage of containers for Stage-2 (38\%) and Stage-3 (36\%). This is because, in addition to stage-aware batching, \emph{Fifer}'s proactive container scaling policy reduces the aggressive reactive scaling of containers. The number of containers is less for Stage-2 (21\%) because its a very short running stage (less than 2\% of total execution time) and thereby results in early scale-in of idle containers. Though aggressive batching can result in SLO violations (15\% and 12\% more than \emph{Bline} for \emph{SBatch} and \emph{RScale} respectively), \emph{Fifer} ensures similar SLO violation as in \emph{Bline}, because the LSTM model can well adapt to variations in arrival rate. \\
\begin{figure}
\begin{minipage}[t]{0.99\linewidth}
\begin{subfigure}[t]{.45\textwidth}
\centering\includegraphics[width=0.99\textwidth]{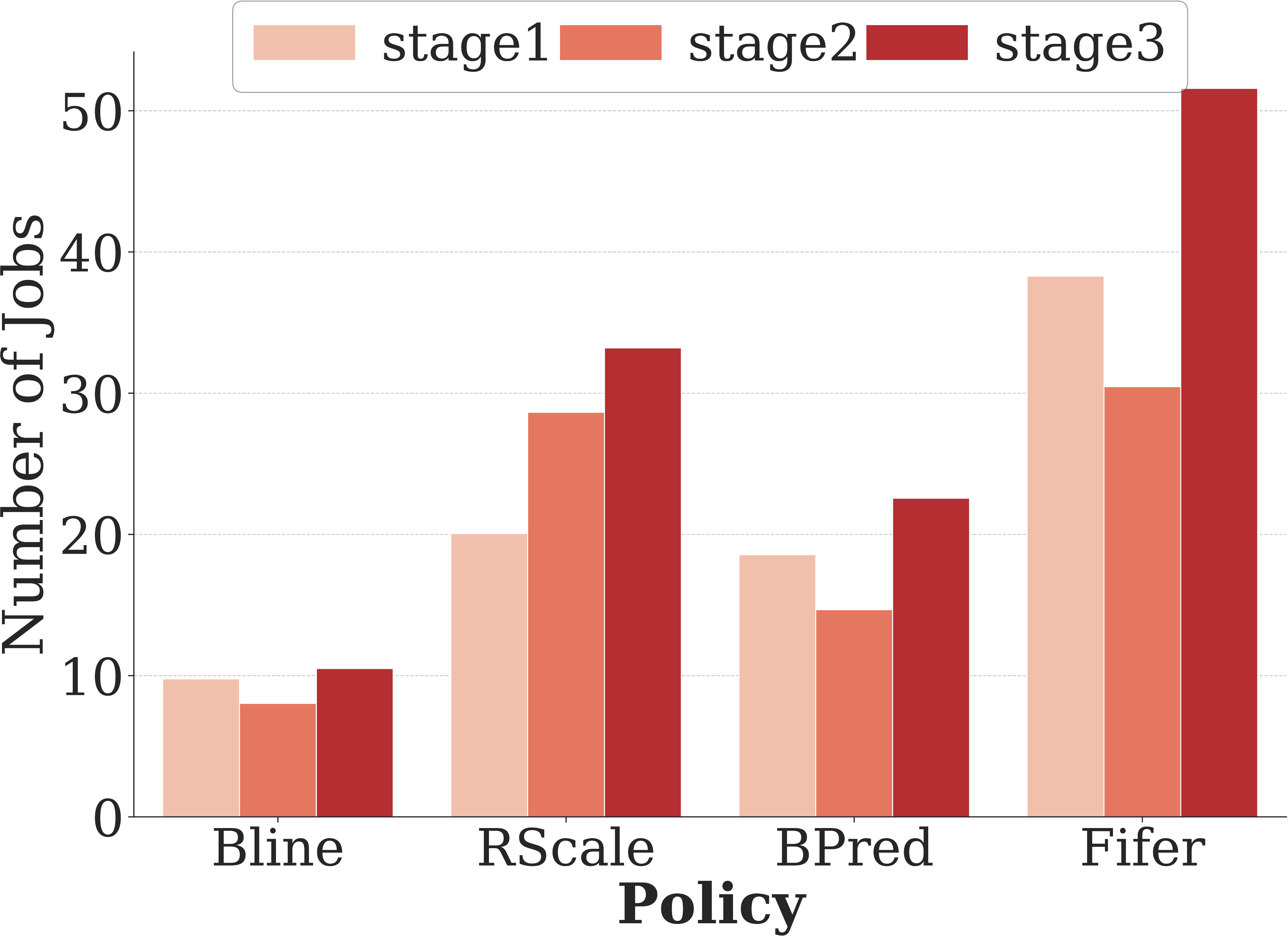}
\caption{Average number of jobs executed per container (JPC)}
\label{fig:cont-util}
\end{subfigure}
\begin{subfigure}[t]{0.50\textwidth}
\centering
\includegraphics[width=0.99\textwidth]{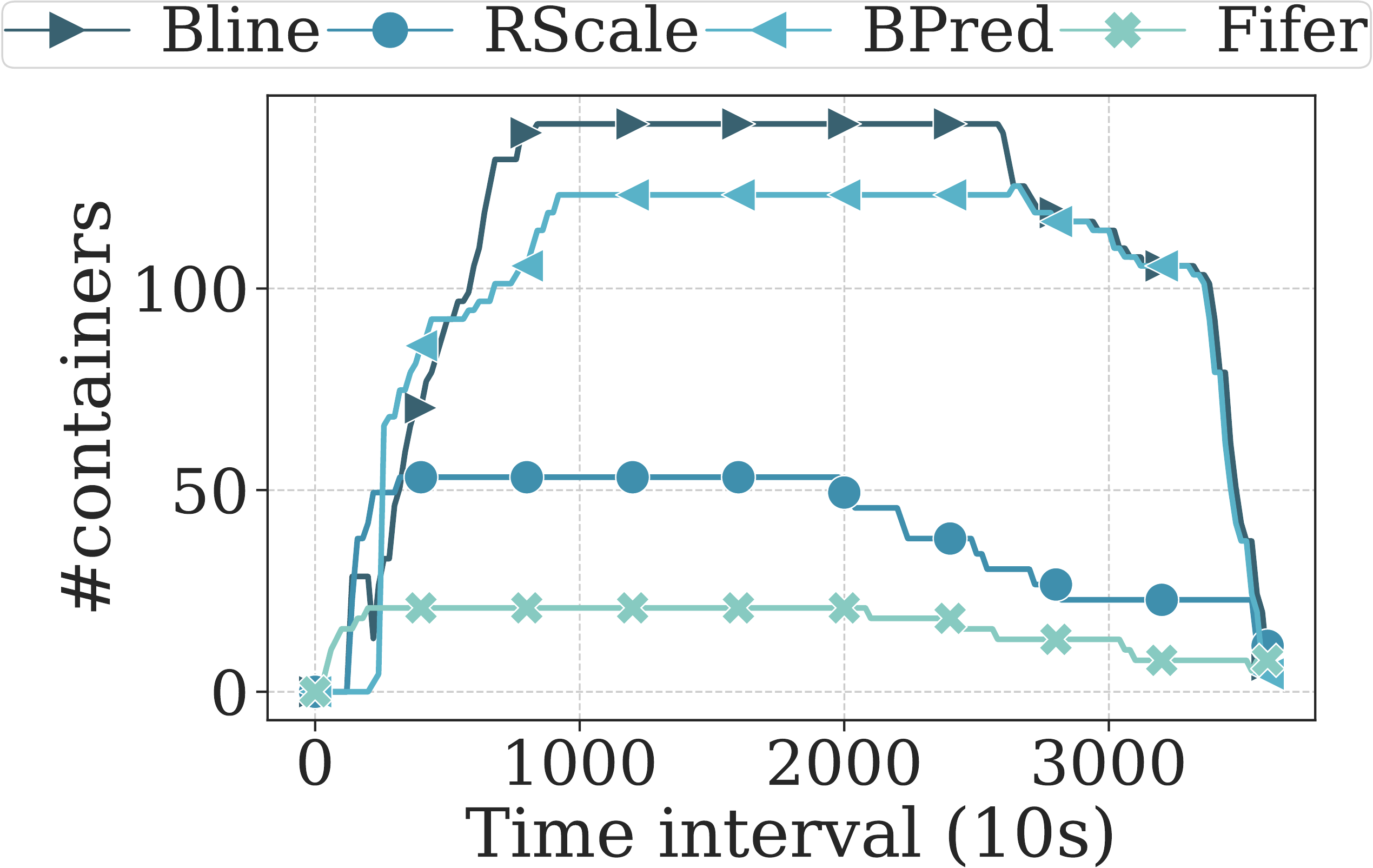}
\caption{Number of Containers spawned over time.}
\label{fig:cont-time}
\end{subfigure}
\end{minipage}
\caption{Sources of Improvement.}
\end{figure}
\noindent{\textbf{Container Utilization:}} 
Figure~\ref{fig:cont-util} plots the average number of tasks (requests) executed by a container in all stages. We define container utilization as Requests executed per Container (RPC). It is evident that \emph{Fifer} has the maximum RPC across all stages. Intuitively, for a given total number of requests, higher RPC indicates that a lesser number of containers are being spawned. It can be seen that both \emph{Bline} and \emph{BPred} scheme always spawn a large number of containers due to non-batching nature, which is exacerbated, especially for short running stage-2 (RPC of 8.03\% and 11.67\%). Though both \emph{RScale} and \emph{Fifer} employ request-batching. \emph{Fifer} still has 12.6\% better RPC on average across all stages than \emph{RScale}. This is because \emph{Fifer} inherently minimizes over-provisioned containers as a result of proactive container spawning.

To better understand the benefits of proactive provisioning, Figure~\ref{fig:cont-time} plots the overall the number of containers spawned measured over intervals of 10s for all four RMs. It can be seen that both \emph{RScale} and \emph{Fifer} adapt well to the request rate distribution, and due to batching they spawn up to  60\% and 82\% fewer containers on average when compared to \emph{Bline} RM. \emph{Fifer} is still 22\% better than \emph{RScale} because \emph{Fifer} can accurately estimate the number of containers required in each stage by using the LSTM prediction. 

\subsubsection{Cluster Energy Savings:} Since in \emph{Fifer} we effectively bin-pack containers to least-resources-available servers, it results in server. \begin{wrapfigure}{r}{0.25\textwidth}
\captionsetup{justification=justified}
    \centering
    \includegraphics[width=.24\textwidth]{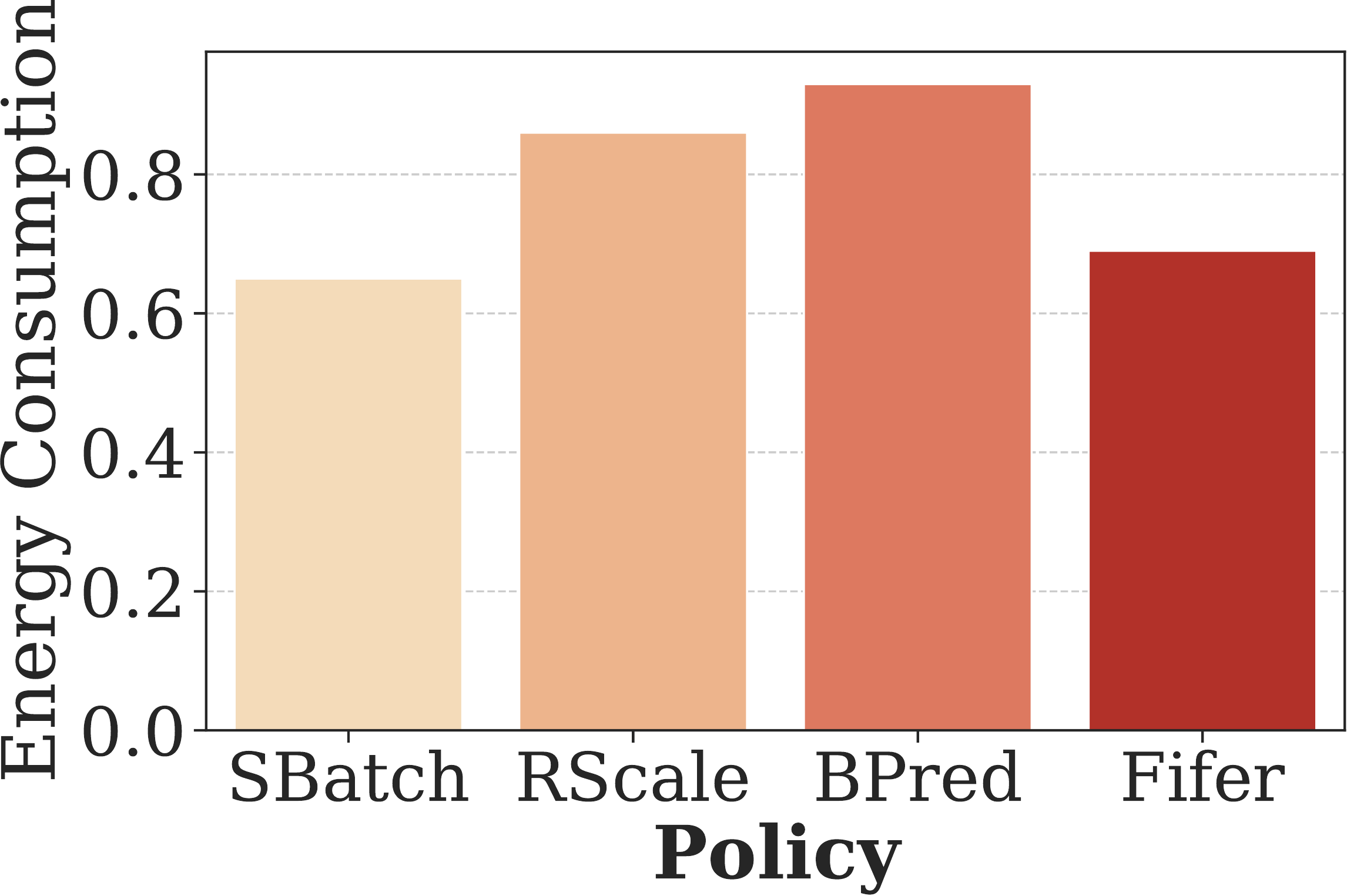}
        \caption{Cluster-wide energy savings normalized to \texttt{Bline}.}
\label{fig:energy}
\end{wrapfigure}consolidation thereby increasing the cluster efficiency. Figure~\ref{fig:energy}
plots the cluster-wide energy as an average of energy consumed across all nodes in the cluster measured over intervals of 10 seconds for the entire workload duration. It can be seen that \emph{Fifer} is 30.75\% more energy efficient than the \emph{Bline} (for heavy workload-mix). This is because \emph{Fifer} can accurately estimate the number of containers at each stage, thereby resulting in all active containers to get consolidated in fewer nodes. \emph{Fifer} is also 17\% more energy efficient than \emph{RScale}, because proactive provisioning reduces the number of  reactively spawned containers. This, in turn, results in reducing the number of idle CPU-cores in the cluster. \emph{Fifer} is almost as energy efficient as \emph{Sbatch} (difference of 4\%), but at the same time, it can scale up/down according to request demand, thus minimizing SLO violations when compared to \emph{Sbatch}. 

\subsubsection{System Overheads:}
As meeting the response latency is one of the primary goals of \emph{Fifer}, we characterize the system-level overheads incurred due to the design choices in \emph{Fifer}. The \emph{mongodb} database is a centralized server which resides on the head-node. We measure the overall average latency incurred due to all reads/writes in the database, which is well within 1.25ms. The LSF scheduling policy takes about 0.35ms on average per scheduling decision. The latency of LSTM request prediction which is not in the critical scheduling path and runs as a background process model is 2.5 ms on average. The time taken to spawn new container, including fetching the image remotely takes about 2s to 9s depending on the size of the container image. 
\begin{figure}
\begin{minipage}[t]{0.99\linewidth}
\begin{subfigure}[t]{.48\textwidth}
\centering
\includegraphics[width=0.99\textwidth]{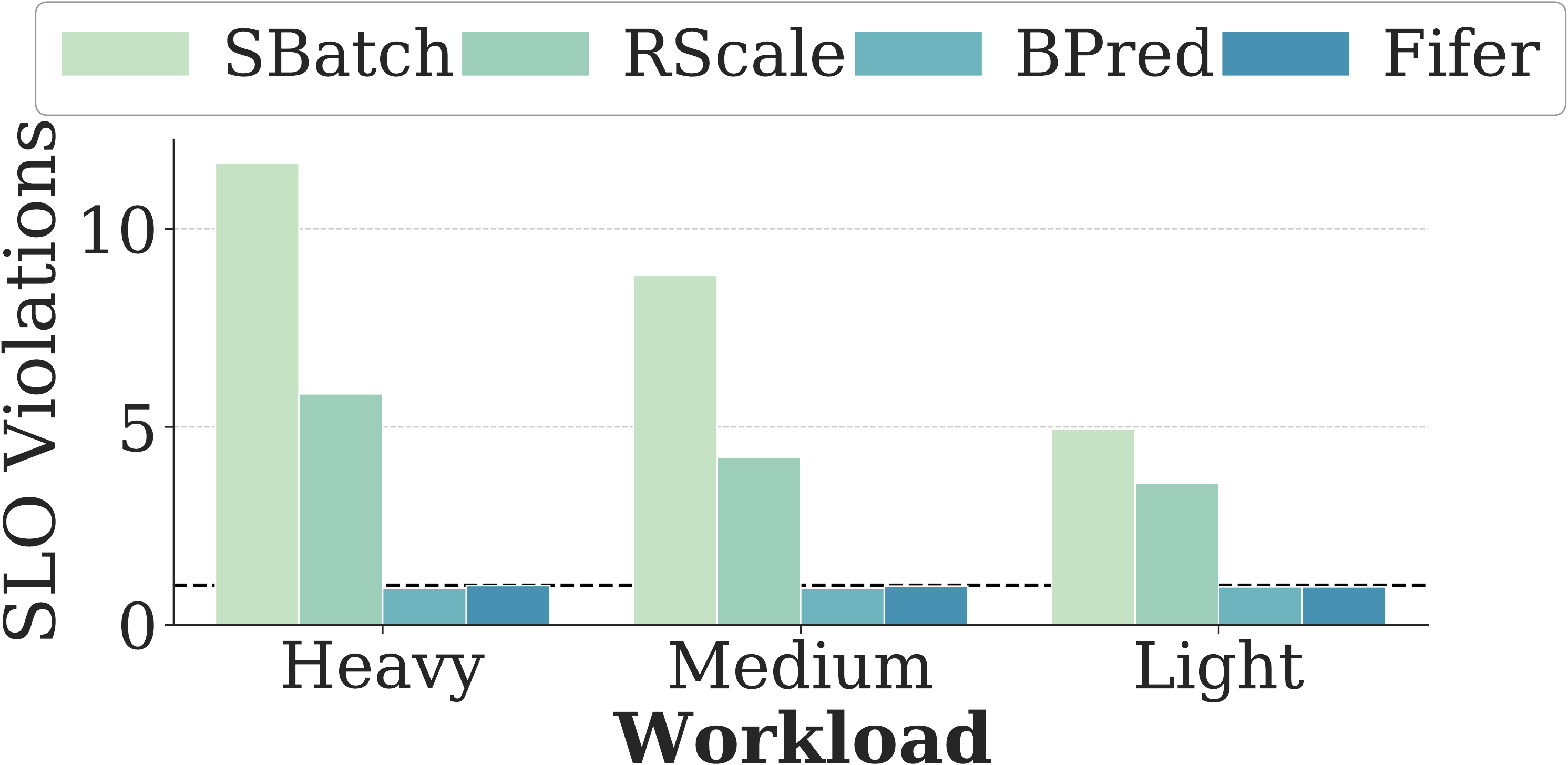}
\caption{SLO violations.}
\label{fig:wiki-sla}
\end{subfigure}
\begin{subfigure}[t]{0.48\textwidth}
\centering
 \includegraphics[width=0.99\textwidth]{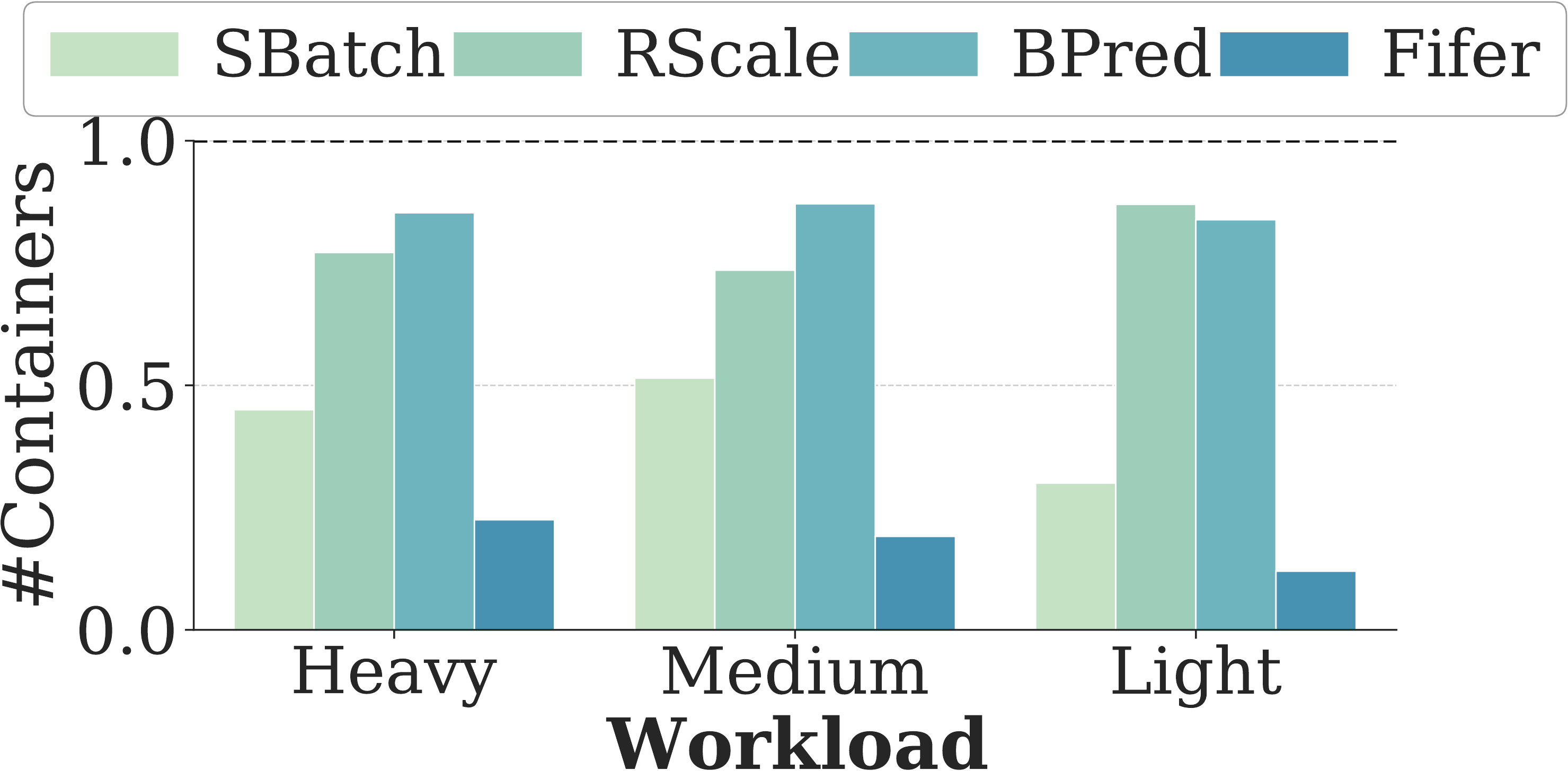}
\caption{Average number of Containers.}
\label{fig:wiki-cont}
\end{subfigure}
\end{minipage}
\caption{Macro Benchmark: Three workloads mix using Wikipedia request arrival trace. Results normalized to \texttt{Bline}.}
\label{fig:wiki}
\end{figure}
\begin{figure}
\begin{minipage}[t]{0.99\linewidth}
\begin{subfigure}[t]{.49\textwidth}
\centering
\includegraphics[width=0.99\textwidth]{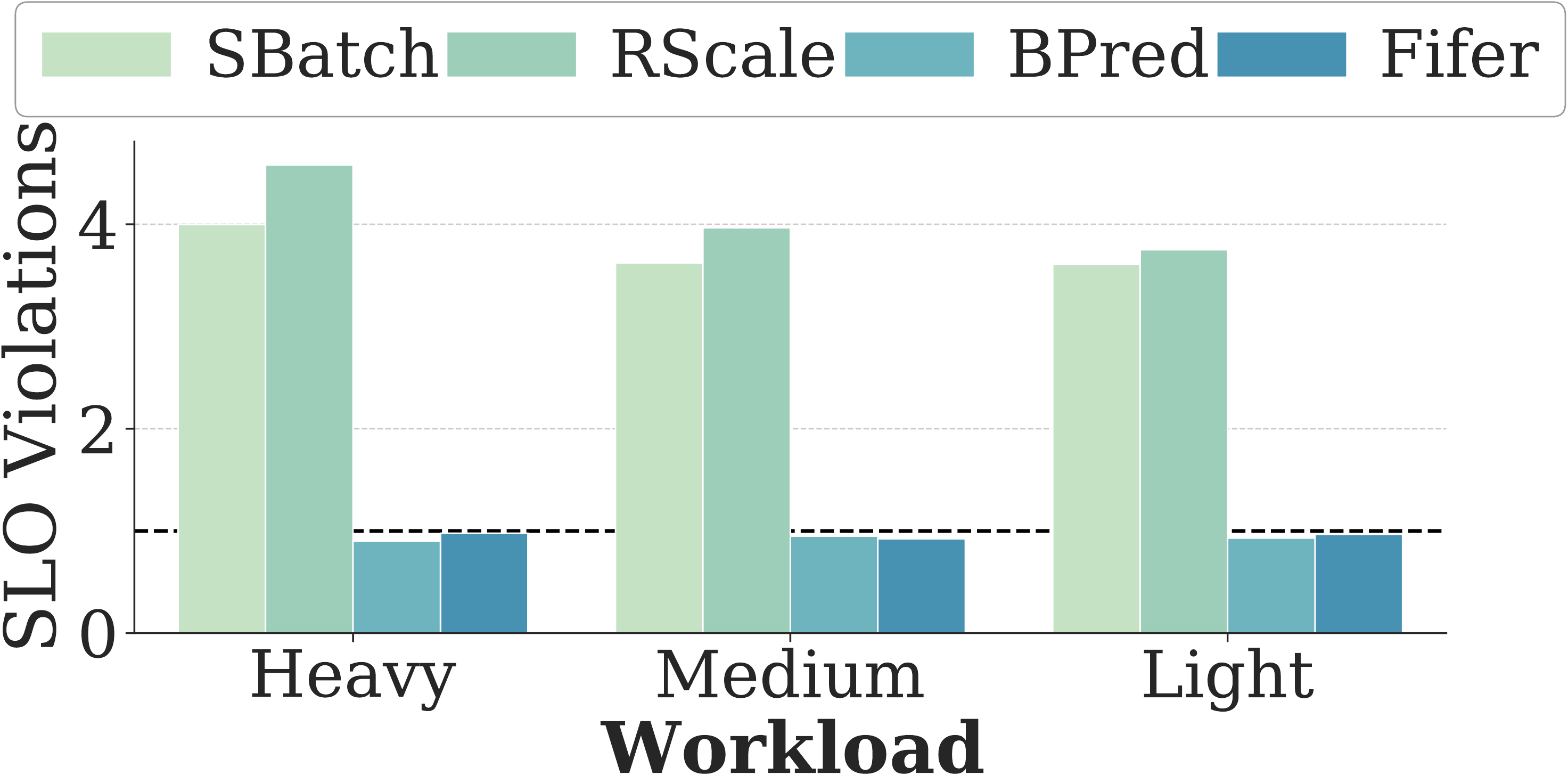}
\caption{SLO violations.}
\label{fig:wits-sla}
\end{subfigure}
\begin{subfigure}[t]{0.49\textwidth}
\centering
 \includegraphics[width=0.99\textwidth]{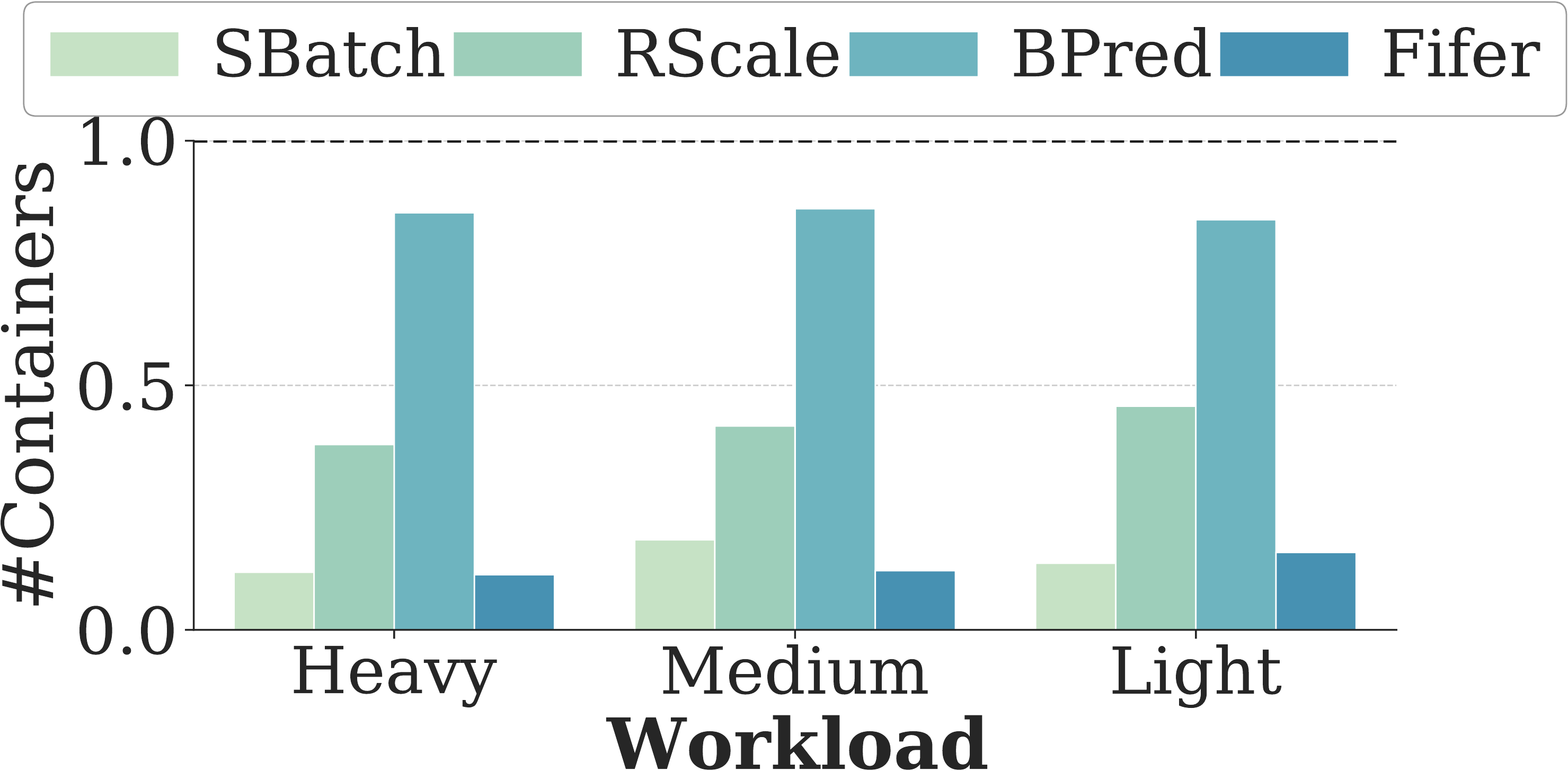}
\caption{Average number of Containers.}
\label{fig:wits-cont}
\end{subfigure}
\end{minipage}
\caption{Macro Benchmark: Three workloads mix using WITS request arrival trace. Results are normalized to Baseline.\vspace{1mm}}
\label{fig:wits}
\end{figure}
\subsection{\textbf{Trace driven Simulation}}
We evaluate our simulator using large scale Wiki and WITS traces. Figure~\ref{fig:wiki} plots the percentage of SLO violations,  and the average number of containers spawned normalized to \emph{Bline}, for the Wiki trace for all three workloads~\footnote{All the results in this subsection are reported as an average for the three workloads-mix.}. The benefits of \emph{Fifer} are significantly higher than those observed  on the real-system. \begin{wraptable}{L}{0.27\textwidth}
\centering
\footnotesize
\begin{tabular}{|l|l|l|l|l|}
\hline
\multirow{2}{*}{\textbf{Policy}} & \multicolumn{2}{l|}{\textbf{Wiki}} & \multicolumn{2}{l|}{\textbf{WITS}} \\ \cline{2-5} 
 & \textbf{Med} & \textbf{Tail} & \textbf{Med} & \textbf{Tail} \\ \hline
\textit{Bline} & 233 & 3967 & 237 & 5807 \\ \hline
\textit{SBatch} & 458 & 13349 & 437 & 17736 \\ \hline
\textit{RScale} & 251 & 10245 & 252 & 12164 \\ \hline
\textit{BPred} & 281 & 4240 & 290 & 5914 \\ \hline
\textit{Fifer} & 413 & 4952 & 354 & 6151 \\ \hline
\end{tabular}
\caption{Median and tail latencies (in milliseconds) for the Wiki and WITS traces using the heavy workload-mix.\vspace{-1mm}}
\label{tab:latency}
\end{wraptable}This is because the Wiki trace follows a diurnal pattern with a highly dynamic load, thus leading to many unprecedented request scale-out and consequently requires more containers to be spawned. Since the \emph{Bline}, \emph{Bpred} and \emph{RScale} RMs employ reactive scaling, they experience higher average number of containers spawned (shown in Figure~\ref{fig:wiki-cont}). Especially, the \emph{RScale} and \emph{Bpred} RM spawns up to 3.5$\times$ more containers on average compared to \emph{Fifer}, still leading to 5\% more SLO violations than \emph{Fifer} (shown in Figure~\ref{fig:wiki-sla}). This is because it cannot predict the variations in the input load. \emph{Fifer}, on the other hand, is resilient to load fluctuations as it utilizes an LSTM-based load prediction model that can accurately capture the variations and proactively spawn containers. The tail latencies are also extremely high for the \emph{RScale} RM, due to the congestion of request queues resulting from cold-starts (shown in Table~\ref{tab:latency}). However, the median latencies follow similar increasing trends, as observed in the real-system.  

Figure~\ref{fig:wits} plots the percentage of SLO violations and average containers spawned normalized to \emph{Bline}, for WITS trace for all three workloads. The WITS trace exhibits sudden peaks due to a higher peak-to-median ratio in arrival rates (the peak (1200 req/s) is 5$\times$ higher than the median (240 req/s)). This sudden surge leads to very high tail latencies (shown in Table~\ref{tab:latency}). \emph{Fifer} can still reduce tail-latencies by up to 66\% when compared to \emph{Sbatch} and \emph{RScale} policies. The amount of SLO violations (shown in Figure~\ref{fig:wits-sla}) are considerably lower for all policies in comparison to Wiki trace, due to less dynamism in the arrival load. 
However, \emph{Fifer} still spawns 7.7$\times$, 2.7$\times$ fewer  containers on average (Figure~\ref{fig:wits-cont}), when compared to the \emph{\emph{Bpred}} and \emph{RScale} RMs, respectively. The savings of \emph{Fifer} with respect to \emph{RScale} are lower when compared to WIKI trace, because the need to spawn additional containers by reactive scaling is considerably reduced when there less frequent variations in arrival rates. Similar to the real-system, \emph{Fifer} ensures SLO's to the same degree (up to 98\%) as \emph{Bline} and Bpred RMs. 
\subsubsection{Effect of Coldstarts: }
\begin{wrapfigure}{r}{0.26\textwidth}
\captionsetup{justification=justified}
    \centering
    \includegraphics[width=.25\textwidth]{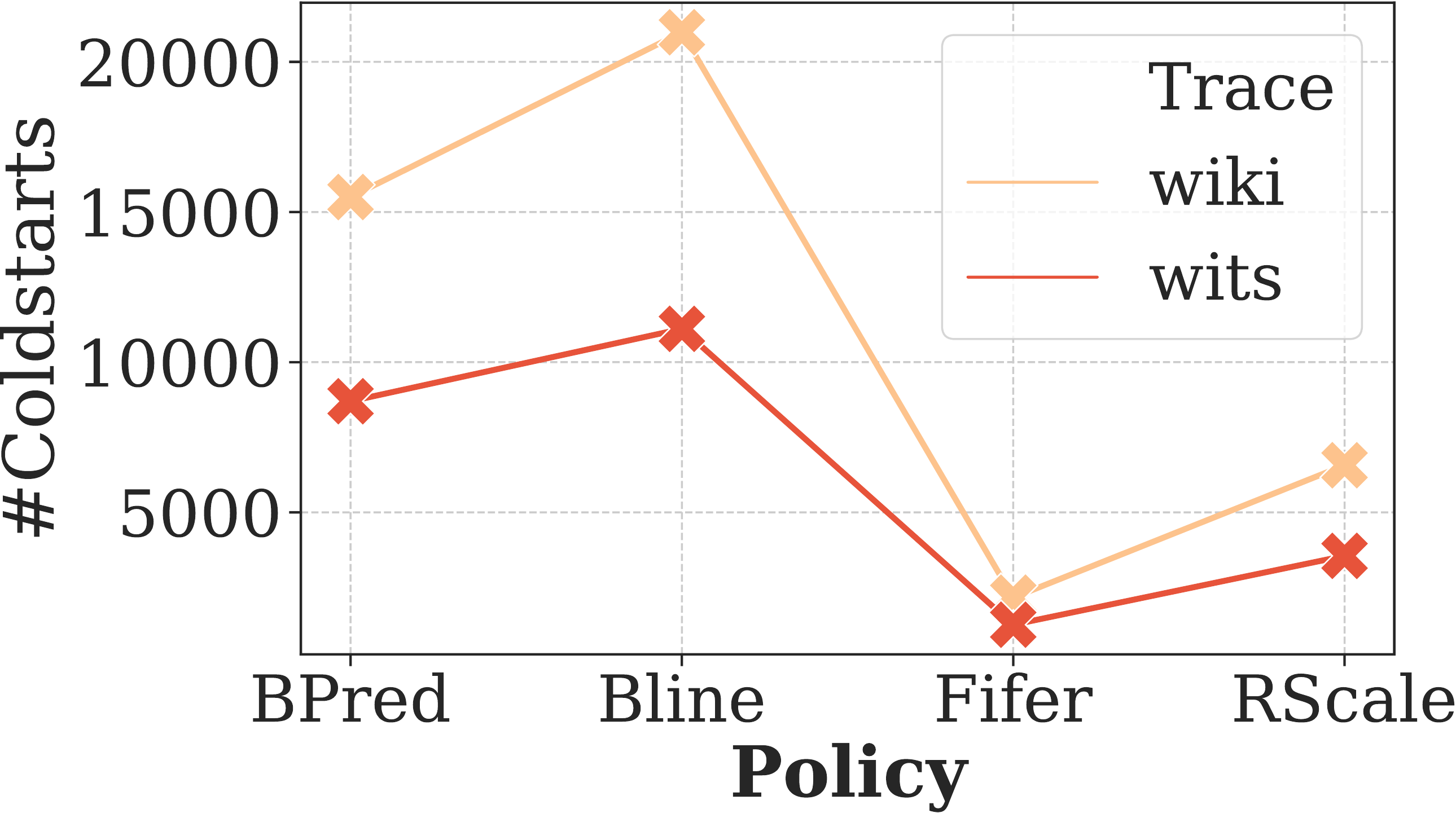}
    \vspace{-2mm}
    \caption{Number of Coldstarts.}
\label{fig:cold-start}
\end{wrapfigure}

Figure~\ref{fig:cold-start} plots the number of cold-starts incurred by three different RMs for a 2 hour snapshot of both traces. 
 It can be seen that, \emph{Fifer} reduces the number of cold-starts by up to 7$\times$ and 3.5$\times$, when compared to Bpred for the Wiki and Wits trace, respectively. Though \emph{RScale} also significantly reduces cold-start when compared to \emph{Bline} and \emph{BPred}, \emph{Fifer} is still 3$\times$ better than \emph{RScale}, This is because \emph{Fifer} avoids a large number of reactive cold-starts by accurately spawning containers in advance. It should also be pointed out that the number of cold-starts are more for WIKI trace because the average request rate is 5$\times$ higher than WITS trace.

\section{Related Work} 
\label{sec:related}
\textbf{Managing Microservice Chains:} 
The most relevant and recent prior works to ours that have looked into resource management for microservice chains can be summarized as follows: (i) \emph{Grandslam}~\cite{grandslam} proposes a dynamic slack-aware scheduling policy for guaranteeing SLOs in shared microservice execution frameworks. Unlike Grandslam, \emph{Fifer} focuses on entirely different goals of container provisioning combined with container scalability, in the context of RMs used in serverless computing frameworks. As demonstrated by our results, Grandslam (\emph{RScale} policy) suffers from SLO violations, while scaling containers due to dynamic load variations.  (ii) \emph{Archipelago}~\cite{archipelago} aims to provide low latency scheduling for both monolithic and chained microservices. In contrast, \emph{Fifer} dives deep in its goal, such that it aims to specifically optimize for microservice-chains with the primary objective of increasing resource utilization by minimizing \#containers and \#servers used, without compromising on SLOs.  Table~\ref{tbl:related} provides a comprehensive analysis of all the features of Fifer, comparing it with other relevant works. \\
\begin{table}
\footnotesize
\begin{center}
\resizebox{0.45\textwidth}{!}{%
 \begin{tabular}{||c | c | c | c |c | c | c | c|c||} 
 \hline
 \textbf{Features} & \begin{turn}{90}Grandslam \cite{grandslam}\end{turn} & \begin{turn}{90}Power-chief \cite{yang2017powerchief}\end{turn} & \begin{turn}{90}Time-Trader \cite{timetrader}\end{turn} &
 \begin{turn}{90}Parties
 \cite{parties}\end{turn} &
 \begin{turn}{90}MArk \cite{mark}\end{turn} & \begin{turn}{90}Archipelago \cite{archipelago}\end{turn}&
 \begin{turn}{90}Swayam \cite{swayam}\end{turn}& \begin{turn}{90}\emph{Fifer} \end{turn} \\
 \hline
 \textbf{Server consolidation} & \cmark & \cmark & \cmark & \xmark & \cmark & \xmark & {\cmark} & {\cmark}\\  
 \hline
  \textbf{SLO Guarantees} & \cmark & \xmark & \cmark & \cmark & \cmark & \cmark & \cmark & {\cmark}\\
 \hline
 \textbf{Function Chains} & \cmark & \cmark & \xmark & \xmark &\xmark & \cmark & \xmark & {\cmark} \\
  \hline
 \textbf{Slack based scheduling} & \cmark & \cmark & \cmark &\cmark & \xmark & \cmark & \xmark & {\cmark}\\
   \hline
\textbf{Slack aware batching} & \cmark & \xmark & \xmark &\xmark & \xmark & \xmark & \xmark & {\cmark}\\
\hline
 \textbf{Energy Efficient} & \xmark & \cmark & \cmark & \xmark &\xmark & \xmark & \cmark & {\cmark}\\

 \hline
 \textbf{Autoscaling Containers} & \xmark & \cmark & \xmark & \xmark &\cmark & \cmark & \cmark & {\cmark}\\
 \hline
 \textbf{Request Arrival prediction} & \xmark & \xmark & \xmark & \xmark &\xmark & \cmark & \cmark & {\cmark}\\
\hline
\end{tabular}}
\end{center}
  \caption{{\color{black}Comparing the features of \emph{Fifer} with other state-of-the-art resource management frameworks.}}
  \label{tbl:related}
\end{table}
\textbf{Scheduling and Resource Management:}\\
$\bullet$ \textit{Private Cloud}:A large body of work~\cite{ousterhout2013sparrow,Delimitrou:tarcil,delgado2015hawk,delgado2016job,karanasos2015mercury} have looked at ensuring QoS guarantees for latency critical applications by developing sub-millisecond scheduling strategies using both distributed and hybrid schedulers. Some works~\cite{knots,proteus,baymax} employ prediction-based scheduling in RMs for executing latency-critical tasks that are co-located with batch tasks. However, none of these techniques caters to the needs of efficiently executing applications with microservice chains, as they all look at microservices similar to conventional monolithic applications.\\ 
$\bullet$ \textit{Public Cloud}: There are several research works that optimize for the resource provisioning cost in the public cloud. These works broadly fall into two categories: (i) tuning the auto-scaling policy based on changing needs (e.g., Spot, On-Demand)~\cite{swayam,tributary,proteus,stratus,Wang:2017:UBI:3107080.3084448,buRScale,exosphere}, 
(ii) predicting peak loads and offering proactive provisioning based auto-scaling policy~\cite{swayam,mark,10.5555/2748143.2748357,spotcheck,tributary}. \emph{Fifer} uses similar load prediction models and auto-scales containers but with respect to serverless function chains. Swayam~\cite{swayam} is relatively similar to our work such that, it handles  container provisioning along with load-balancing. Unlike \emph{Fifer} which looks at micro-service chains, Swayam is specifically catered for single-function machine learning inference services.\\
\textbf{Exploiting Slack:}
Exploiting slack between tasks is a well-known technique, which has been applied in various domains of scheduling, including SSD  controllers~\cite{ssd1,ssd2}, memory controllers~\cite{memory1,memory2}, and network-on-chip~\cite{mem1,mem2}. In contrast to exploiting slack, we believe the novelty aspect lies in identifying the slack in relevance to the problem domain and designing policies to utilize the slack effectively. \\
\textbf{Mitigating Cold-starts:}
Many recent works~\cite{215949,akkus2018sand,sock} propose optimizations to reduce container setup overheads. For example, SOCK~\cite{sock} and SAND~\cite{akkus2018sand} explore optimizations to reduce language framework-level overheads and the step function transition overheads, respectively. Some works propose to entirely replace containers with new virtualization techniques like Firecracker~\cite{firecracker} and uni-kernels~\cite{unikernel}. Complementary to these approaches, \emph{Fifer} tries to decouple container cold-starts from request execution time.
\section{Discussion and Future Work} 
\label{sec:discussion}
\textbf{Design Limitations:} We set SLO to be within 1000ms, which is the typical user-perceived latency. Note that changing the SLO would result in different slacks for application stages. While providing execution time and SLO information is an offline step in \emph{Fifer}, for longer running applications where execution time is greater than 50\% of SLO, the benefits of batching would be significantly reduced. 

Our execution time estimates are limited to ML-based applications, but our schemes can be applied to all other applications which have predictable execution times. 
Also, the applications we consider are linearly chained without any dynamic branches. We leave the exploration of dynamic microservice chains for future work.

All decisions related to container scaling, scheduling and load-prediction are reliant on the centralized database which can become a potential bottleneck for a large scale system with thousands of nodes. This can be mitigated by distributing the database to a subset of nodes or by using sophisticated solutions like Zookeeper~\cite{zookeeper}. The LSTM model in \emph{Fifer} is pre-trained using 60\% of the arrival trace. Although both the Wiki and WITS capture typical arrival rate scenarios in a datacenter, the model will fail to accurately predict for a completely unforeseen arrival pattern. In such cases, the LSTM model parameters can be constantly updated by retraining in the background with new arrival rates. We leave both these discussions to future work. 
\\
\textbf{Cloud Provider Support:} The cold-start measurements and characterizations  in \emph{Fifer} are mainly based on AWS. However, the main design of \emph{Fifer} can be extended in theory to other major cloud platforms as well. We also rely on the serverless platform provider to expose API's for the tenants to specify their application SLO requirements which are crucial for estimating the slack. Such an API would better enable the provider to auto-configure tenants’ execution environments that would be invaluable in improving resource efficiency~\cite{234821}.
\section{Concluding Remarks}
\label{sec:conclusion}
There is wide-spread prominence in the adoption of serverless functions for executing microservice-based applications in the cloud. This introduces critical inefficiencies in terms of scheduling and resource management for the cloud provider, especially when deploying a large number millisecond-scale latency-critical functions. In this paper, we propose and evaluate \emph{Fifer}, a stage-aware adaptive resource management framework for efficiently running function-chains on serverless platforms by ensuring high container utilization and cluster efficiency without compromising on SLOs. \emph{Fifer} makes use of a novel combination of stage-wise slack awareness along with proactive container allocations using an LSTM-based load prediction model. The proposed technique can intelligently scale-out and balance containers for every individual stage. Our experimental analysis on an 80 compute-core cluster and large scale simulations show that \textit{Fifer} spawns up to 80$\%$ fewer containers on average, thereby improving container utilization by 4$\times$ and cluster efficiency by 31\%. 


\bibliographystyle{ACM-Reference-Format}
\bibliography{references}


\end{document}